\documentclass[11pt,a4paper]{article}
\pdfoutput=1
\usepackage{jheppub}
\usepackage{amsmath,amscd}
\usepackage{amsfonts}
\usepackage{amssymb}
\usepackage{graphicx}
\usepackage{color}
\usepackage[normalem]{ulem}
\usepackage{hyperref}
\usepackage{enumerate}
\usepackage{mathtools}
\newcommand{\RR}{\mathbb{R}} % Reali
\newcommand{\ZZ}{\mathbb{Z}} % Interi
\newcommand{\G}{\mathcal{G}}

\def\tr         {{\rm  tr}}
\def\cala         {{\cal A}}

\def\calc         {{\cal C}}

\def\calh         {{\cal H}}

\def\calm         {{\cal M}}
\def\caln         {{\cal N}}

\def\calr         {{\cal R}}
\def\cals         {{\cal S}}
\def\calt         {{\cal T}}

\def\be{\begin{equation}}
\def\ee{\end{equation}}
\def\bea{\begin{eqnarray}}
\def\eea{\end{eqnarray}}

\def\a{\alpha}
\def\b{\beta}
\def\h{\eta}
\def\g{\gamma}
\def\G{\Gamma}
\def\d{\delta}

\def\o{\omega}
\def\O{\Omega}
\def\p{\pi}
\def\r{\rho}
\def\z{\zeta}
\def\s{\sigma}
\def\S{\Sigma}
\def\t{\tau}

\def\sF{{{ F}\!\!\!\!\hskip.8pt\hbox{\raise1pt\hbox{/}}\,}}
\def\som{{{ \omega}\!\!\!\!\hskip.8pt\hbox{\raise1pt\hbox{/}}\,}}
\def\sJ{{{\rm J}\!\!\!\!\hskip.8pt\hbox{\raise1pt\hbox{/}}\,}}

\newcommand\eqqm{\stackrel{\mathclap{\normalfont\mbox{?}}}{=}}

\def\pa{\partial}

\def\to{\rightarrow}
\def\nonu{\nonumber \\{}}
\def\half{{1 \over 2}}

\def\md{{\rm mod\ }}

\usepackage{cancel}
\usepackage{subcaption} %for subfigures

\usepackage{amsthm}
\theoremstyle{plain}

\theoremstyle{plain}

\theoremstyle{plain}
\newtheorem*{prop*}{\protect\propositionname}

\providecommand{\lemmaname}{Lemma}
\providecommand{\propositionname}{Proposition}
\providecommand{\theoremname}{Theorem}
\title{%On the role of  surface defects in  ensemble holography}
	Wormholes and surface defects in rational ensemble holography}
\author{Joris Raeymaekers and Paolo Rossi}

\affiliation{CEICO, Institute of Physics of the Czech Academy of Sciences,\\  Na Slovance 2, 182 21 Prague 8, Czech Republic.}
\emailAdd{(joris,rossip)@fzu.cz} 

\abstract{We study wormhole contributions to the bulk path  integral in holographic models which are dual to    ensembles of rational free boson conformal field theories. We focus on  the path integral on a geometry connecting two toroidal boundaries, which should capture the variance of the ensemble distribution. 
	We show  that this requirement leads to a nontrivial set of constraints which generically
	picks out the uniform, maximum entropy, ensemble distribution. Furthermore, we show that  the two-boundary path integral should receive contributions from `exotic' wormholes, which arise from the inclusion of topological surface defects.}

\begin{document}
 \maketitle

\section{Introduction and summary}
Recent developments have led to a revived interest in the role of wormhole geometries contributing to the Euclidean path integral in gravity. On one hand, inclusion of replica wormholes in the path integral for the entanglement entropy of black hole radiation have shed light on the information puzzle \cite{Penington:2019kki,Almheiri:2019qdq}.   On the other hand, contributions from wormholes connecting different asymptotic boundaries  lead to non-factorizing observables and are a sign that the path integral computes an average over an ensemble of CFTs.

Perhaps the most clear-cut instance  of such  ensemble holography is that of JT gravity, where the contribution of the `double-trumpet' wormhole and its generalizations are directly related to the fact that  the model is dual to an ensemble of quantum mechanical theories \cite{Saad:2019lba}. These insights from two-dimensional gravity have also led to a tentative reinterpretation of Maloney and Witten's  result \cite{Maloney:2007ud} for the partition function of pure three dimensional anti-de Sitter gravity, which displays features of an ensemble average. For one, the resulting continuous spectrum is most naturally interpreted as describing an average of theories with discrete spectra. Furthermore, Cotler and Jensen found an explicit bulk wormhole contributing to the two-boundary partition function \cite{Cotler:2020ugk}.   
The putative ensemble interpretation of the Maloney-Witten result for the gravity path integral on a solid torus with boundary complex structure $\t$ would imply  
an  
identity of the form
\be 
%Z^{\rm bulk} [\t] =
 \sum_{\g \in \G / \G_f} \left|\chi_{\rm 0} (\g \t )\right|^2 \eqqm \langle Z  [\t] \rangle\label{MWaverage}
\ee
Here, the left hand side is a so-called Poincar\'e sum with  $\chi_{\rm 0}$  Virasoro vacuum character, $\G = PSL(2,\ZZ)$ is the modular group and $\G_f$ is the subgroup leaving $ |\chi_{0} |^2$ invariant. The expectation value of the CFT partition function $Z[\t ]$ on the right-hand side stands for an ensemble average over all Virasoro CFTs. Since we have at present no control over this ensemble (nor how to define an integration measure to average over it), making (\ref{MWaverage}) more precise remains  a daunting task\footnote{Furthermore, the Poincar\'{e} sum defining the left hand side also has problematic features  \cite{Keller:2014xba,Benjamin:2019stq}, see  \cite{Benjamin:2020mfz,Maxfield:2020ale} for proposed resolutions.}.

It is encouraging however that relations of the type   (\ref{MWaverage}) do  turn out to hold in situations with  enhanced symmetry.  That is, if we replace $\chi_{\rm 0}$ by the vacuum character of an extended chiral algebra for which we have control over the moduli space of CFTs,  (\ref{MWaverage}) can often be turned into a mathematical identity. The candidate bulk theory giving rise to  (\ref{MWaverage}) is then  an `exotic Chern-Simons gravity', where one starts from a Chern-Simons theory realizing the appropriate chiral algebra, and supplements it with a prescription to  sum over certain  topologies in the path integral. 

  A prime example is the case where $\chi_{0}$ is taken to be the vacuum character of a $u(1)^D$ current algebra \cite{Afkhami-Jeddi:2020ezh,Maloney:2020nni}. Here, the Siegel-Weyl formula is an identity of the form  (\ref{MWaverage}) where the right hand  side  is an average over  Narain moduli space of free boson CFTs with respect to a natural measure. Further studies and extensions of  Narain holography include \cite{Perez:2020klz,Dymarsky:2020pzc,Datta:2021ftn,Benjamin:2021wzr,Ashwinkumar:2021kav,Dong:2021wot,Collier:2021rsn,Benjamin:2021ygh,Dymarsky:2021xfc,Chakraborty:2021gzh,Henriksson:2022dnu,Kawabata:2022jxt,Kames-King:2023fpa,Alam:2023qac,Kawabata:2023usr,Kawabata:2023iss,Aharony:2023zit,Ashwinkumar:2023ctt}. 

As we already mentioned, the lack of factorization of ensemble-averaged  two-boundary CFT partition function can be interpreted in the bulk in terms of wormhole contributions to the path integral where the boundaries are connected through the bulk. The wormhole contribution  $Z_{\rm wh} [\t_1,\t_2]$ is the connected part of the expectation value of the product of two partition functions 
\bea 
Z_{\rm conn}^{(2)} [\t_1,\t_2] &=& \langle Z [\t_1] Z [\t_2] \rangle - \langle Z [\t_1]\rangle \langle  Z [\t_2]\rangle \nonu
&=& \langle\left( Z [\t_1] -\langle Z  [\t_1] \rangle \right) 
\left( Z [\t_2] -\langle Z  [\t_2] \rangle \right) \rangle , \label{whPFintro}
\eea
 and, as rewritten in the second line, measures the variance of the ensemble distribution.
The bulk interpretation of this expression however leads to a puzzle which was 
 pointed out in \cite{Collier:2021rsn} and which inspired our  investigations. The authors of \cite{Collier:2021rsn} computed the right hand side of (\ref{whPFintro}) in the Narain ensemble and compared it to Cotler and Jensen's direct computation \cite{Cotler:2020ugk} from the path integral on a wormhole geometry.  
 While this direct computation captures part of the result, other terms in the ensemble expression are not accounted for. It therefore appears that the bulk dual to the ensemble requires extra contributions from as yet unknown sectors which we will refer to as `exotic wormholes'.

Our goal in this note is to  clarify the holographic interpretation of the two-boundary partition function, including a version of the exotic wormhole puzzle,
 in a simpler setting where the dual ensemble consists of rational CFTs. The study of identities of the type (\ref{MWaverage}) in generic rational CFTs was recently undertaken in \cite{Meruliya:2021utr} and, in the case of ensembles of Virasoro minimal models, goes back to   \cite{Castro:2011zq}. Further studies of rational ensemble holography include \cite{Meruliya:2021lul,Romaidis:2023zpx,Henriksson:2021qkt,Buican:2021uyp, Benini:2022hzx,Henriksson:2022dml}. 
 These models  provide discrete analogs\footnote{While in chaotic CFTs, it is widely believed that the ensemble average captures  self-averaging behaviour of certain observables in a single CFT, such an interpretation will not apply to the integrable rational CFTs under study.}   to Narain holography, since both sides of the relation (\ref{MWaverage}) contain only a finite number of terms. In particular, the right hand side of (\ref{MWaverage}) becomes a sum over a finite set of modular invariant partition functions
\be 
\langle Z  [\t] \rangle = \sum_{I = 1}^\calm  \r_I  Z _I [\t].\label{ensembleweightsintro}
\ee
Although the ensemble weights $\r^I$ can be computed from eq. (\ref{MWaverage}) on a case-by-case basis, a general expression for them in terms of the CFT data  is currently not known. 

In this note we will restrict attention to % study the wormhole puzzle outlined above in 
what is arguably the simplest class of rational CFT ensembles, consisting of free compact boson CFTs where the radius squared is a rational number. %Our focus here will be on ensemble , where the 
These ensembles  may be viewed as  measure zero subsets of the Narain moduli space on which the chiral algebra is enhanced and the CFT becomes rational, 
and were studied by one of us in the context of ensemble holography in \cite{Raeymaekers:2021ypf}. The candidate bulk theory is in this case an exotic Chern-Simons gravity with compact $U(1)_k \times  U(1)_{-k}$ gauge group. It was shown in \cite{Raeymaekers:2021ypf} that, when the integer level $k$ does not contain any square divisors,  the bulk theory describes the maximum entropy  ensemble where the weights $\r_I$ in (\ref{ensembleweightsintro})  are all equal. 

Our main object of study is the connected two-boundary partition function $Z_{\rm conn}^{(2)}$ (\ref{whPFintro}) in this class of models. In the bulk description, %On general grounds  
it  can be expressed as a Poincar\'e sum of the form \cite{Cotler:2020ugk}
\be 
Z_{\rm conn}^{(2)} [\t_1,\t_2] =  \sum_{\g \in  \G / \G_f} Z_{\rm seed}^{(2)} [ \t_1, \g \t_2] \label{zwhbulkintro}
 \ee
where $ Z_{\rm seed}^{(2)}$ is a seed amplitude   representing a functional  integral on  a wormhole geometry connecting two toroidal boundaries with complex structure moduli $\t_1$ and $\t_2$.  As pointed out in \cite{Cotler:2020hgz}, the form   of  $ Z_{\rm seed}^{(2)}$ is constrained due to its three-dimensional  origin. 
Taking these constraints  into account, we point out that matching (\ref{zwhbulkintro}) with its proposed dual (\ref{whPFintro}) leads in general to an overconstrained system, restricting  not only the form of $Z_{\rm seed}^{(2)}$ but also the ensemble weights $\r_I$. When the level $k$ is non-prime and free of square divisors, our analysis  shows that the only consistent solution is in fact the maximum entropy distribution with equal $\r_I$.
This  can be viewed as a nontrivial consistency check on the proposed  ensemble-holographic models with starting point (\ref{MWaverage}), and actually rules out many other putative models where one would  replace the vacuum character $|\chi_0|^2$  in (\ref{MWaverage}) by a more general combination of characters.

We also compare our expression for  $Z_{\rm seed}^{(2)} [ \t_1,  \t_2]$  with a direct bulk computation of {  the modulus square of a} $U(1)_k$ Chern-Simons partition function on a wormhole geometry connecting two toroidal boundaries. %the torus times an interval. 
The latter partition function  $Z^{\rm CS}_{T^2 \times I}(\t_1, \t_2)$ is known from the  work \cite{Elitzur:1989nr}  on the canonical quantization on the annulus and can be viewed as the equivalent of the Cotler-Jensen wormhole partition function \cite{Cotler:2018zff,Cotler:2020hgz} in rational CFTs. As  it turns out,   $Z^{\rm CS}_{T^2 \times I}$   captures only one of the terms %,  corresponding to the  diagonal modular invariant,   
in our expression for the two-boundary seed amplitude $Z_{\rm seed}^{(2)}$. Similar to what happens in Narain holography, we appear to miss a number of contributions from `exotic' wormholes, in this case labelled by nondiagonal modular invariants. However,  in the present context, these have a clear bulk interpretation as  arising from a Chern-Simons % partition function on the torus times an interval
path integral in the presence of a topological surface defect. This follows in straightforward manner from the relation between surface operators in 3D Chern-Simons theory and modular invariants in rational CFT established by Kapustin and Saulina \cite{Kapustin:2010if} (see also \cite{Fuchs:2002cm}).  Therefore, surface defects appear to play a crucial role in ensemble-holographic duality\footnote{See \cite{Choudhury:2021nal,Heckman:2021vzx,Baume:2023kkf} for other instances where topological defects appear in the context of ensemble holography.}.

This paper is organized as follows. In Section \ref{SecRCFTens} we summarize the holographic  paradigm for ensembles of rational  CFTs, and derive general formulas  for the ensemble weights and the  two-boundary wormhole amplitude. In Section \ref{Secratboson} we introduce the ensembles of rational free boson CFTs which will be the focus of our analysis.  In Section \ref{Secsolconstr} we analyze the solutions to the consistency constraints imposed by the  holographic interpretation of the two-boundary wormhole amplitude, both in simple examples and in  general ensembles at square-free level $k$.  Based on these results we point out in Section \ref{Secdefects}   that  ensemble holography requires the inclusion of topological surface defects in the path integral. 
In the Discussion we summarize some lessons to be learned form our findings  and point out   remaining puzzles and possible generalizations.

\section{Rational ensemble holography}\label{SecRCFTens}
In this Section we review some aspects of holography for ensembles of rational CFTs, and derive general expressions for the ensemble weights and the two-boundary wormhole amplitude which will be analyzed in a class of examples in the rest of the paper.

\subsection{Ensembles of rational CFTs} 
In a rational conformal field theory (RCFT), the spectrum   contains only a finite number $\caln$ of irreducible representations of the chiral algebra, which we will denote as $\cala$. We label these representations as $\calr_i, i = 0 , \ldots , \caln-1$, and the corresponding characters %of the representations in this rational CFT 
as $\chi_i (\t), i= 0,\ldots , \caln -1$, with the convention that $\calr_0$ and  $\chi_0$ denote the vacuum representation and the vacuum character. Combining left- and right-moving sectors, RCFTs allow for a finite number   %modular invariant furthermore allow for only a finite number 
$\calm$ of independent modular invariant partition functions which we  label as\footnote{The notation  $[ \t]$ is shorthand for $( \t, \bar \t )$.} $Z_I  [\t], I = 1, \ldots, \calm$.
The coefficients in an expansion  in $\cala \times \overline{\cala}$ characters form  $\caln \times \caln$ matrices  
$M_I$ such that
\be Z_I  [\t] = \chi (\t)^\dagger M_I \chi (\t).\label{ZI}\ee
The matrices $M_I$ are modular invariant in the sense that they commute with the unitary matrices representing the action of the modular group   on the space of characters. 
We will use the convention that $M_1$ is the identity matrix, giving rise to the diagonal modular invariant.  We   restrict our attention to the situation where all modular invariants are physical, meaning that all matrix components of the $M_I$ are integer, and that the vacuum representation appears only once, i.e. 
\be 
M_I^{00} =1.\label{vacunique}
\ee
Furthermore, we   consider here only parity-invariant theories, in which the $M_I$ matrices  are in addition symmetric.

An ensemble of RCFTs with chiral algebra $\cala \times \overline{\cala}$ is  specified by assigning a weight (or probability) $\r_I$ to each of the modular invariant theories. These should be positive and  sum up to one,
\be 
0 \leq \r_I \leq 1, \qquad \sum_{I=1}^\calm \r_I =1.\label{rhonorm}
\ee 
The ensemble-averaged partition function is given by (cfr. (\ref{ensembleweightsintro}))
\be 
\langle Z  [\t] \rangle =  \sum_{I=1}^\calm \r_I  Z_I [\t].\label{ensembleweights}
\ee
One can associate an entropy to the  ensemble distribution,
\be 
S = - \sum_I \r_I \ln \r_I.
\ee
This entropy reaches a maximum  for a  uniform distribution where all CFTs are equally likely,  
\be 
S_{\rm max} = \ln \calm \qquad {\rm for\ } \r_I = \frac{1}{\calm}  \ \forall I.
\ee

Similarly, we can consider the ensemble-averaged $n$-fold  product of partition functions of the form
$\langle \prod_{a=1}^n Z [\t_a] \rangle$.
Of special interest is the connected part of these $n$-point averages. For $n = 2,3$ one finds\footnote{Note that, for the uniform distribution where all ensemble weights are equal,  $\r_I = 1/\calm, \ \forall I$, as will be the case in a large class of examples we will consider, the 2-point function expression further simplifies to
	\be 
Z_{\rm conn}^{(2)}[\t_1,  \t_2] = \frac{1}{\calm^2} \sum_{I <J} \left( Z_I [\t_1]- Z_J [\t_1]\right) \left(Z_I [\t_2]- Z_J [\t_2]\right).
	\ee}
\bea
Z_{\rm conn}^{(2)}[\t_1,  \t_2] &=& \langle  Z [\t_1] Z [\t_2] \rangle - \langle  Z [\t_1] \rangle \langle   Z [\t_2] \rangle \label{variance}\\
&=&  \langle  \left( Z[\t_1] - \langle  Z[\t_1]  \rangle \right)  \left( Z[\t_2] - \langle  Z[\t_2]  \rangle \right)\rangle, \nonu 
Z_{\rm conn}^{(3)}[\t_1,  \t_2, \t_3] &=& \langle  Z [\t_1] Z [\t_2]  Z [\t_3] \rangle - \left( Z_{\rm conn}^{(2)}[\t_1,  \t_2] \langle  Z [\t_3] \rangle + {\rm cyclic}\right) -  \langle  Z [\t_1] \rangle \langle   Z [\t_2] \rangle  \langle   Z [\t_3] \rangle\nonu
&=&  \langle \prod_{a=1}^3 \left( Z[\t_a] - \langle  Z[\t_a]  \rangle \right)\rangle. \nonumber
\eea
More generally, standard generating function methods show that the connected $n$-point partition function measures   $n$-th  cumulant of the ensemble distribution. These can be formally introduced as 
%\footnote{{\color{red}The connected \(n\)-point partition function can be formally defined with the usual procedure of statistics: one introduces a formal \lq generating functional\rq\ \(\mathcal{P}[\lambda]:= \sum_I \rho_I \exp{\left(\int d\tau \lambda(\tau) Z_I[\tau]\right)}\) and a \lq cumulant\rq\ \(\mathcal{W}[\lambda]:=\log(\mathcal{P}[\lambda])\). \(n\)-point partition functions are obtained differentiating \(\mathcal{P}\) with respect to the \lq source\rq\ \(\lambda\), while connected  \(n\)-point functions are obtained differentiating the cumulant \(\mathcal{W}\): \[
%\left. Z^{(n)}_{\text{conn}}[\tau_1,\cdots,\tau_n] = \frac{\delta}{\delta \lambda(\tau_1)}\cdots \frac{\delta}{\delta \lambda(\tau_n) }\mathcal{W}\right|_{\lambda=0}.
%\] The procedure is designed to produce the central moments of the distribution \({\rho_I}\). [\textbf{P: too much?}]}}
\be
Z_{\rm conn}^{(n)}[\t_1, \ldots , \t_n] = \left. \frac{\d}{\d \mathcal{J}[\t_1]}\cdots \frac{\d}{\d \mathcal{J}[\t_n]}\right|_{\mathcal{J}=0} \log \left\langle e^{\int d\t \mathcal{J}[\t] Z[\t]}\right\rangle . \label{Znconn}
\ee

	\subsection{Exotic Chern-Simons gravity in the bulk}
%	For some RCFT ensembles, a candidate bulk description is available in the form of an `exotic Chern-Simons gravity' theory.
Tentative bulk duals to ensembles of RCFTs have been proposed in the form of `exotic Chern-Simons gravities'. These are constructed from  a standard 3D Chern-Simons theory supplemented with a prescription to sum over certain topologies in the path integral, as one would in a  theory of gravity.
	
Through the correspondence between 3D Chern Simons theories and 2D RCFTs \cite{Moore:1989yh,Elitzur:1989nr}, a RCFT with chiral algebra $\cala \times \overline{\cala}$  can be realized in terms of the edge modes   of a Chern-Simons theory on a three-dimensional manifold with boundary, with appropriate boundary conditions. For example, if $\cala$ is the Kac-Moody algebra  $g_k$ associated to a semisimple Lie algebra $g$, we would start from a $G_k \times G_{-k}$ Cherns-Simons theory, while if we want to describe Virasoro or W$_N$ minimal models, we  would start from an $SL(2, \RR) \times SL(2, \RR)$
resp.   $SL(N, \RR) \times SL(N, \RR)$ theory at appropriate values of the level, supplemented with boundary conditions which implement a Drinfeld-Sokolov reduction of the chiral algebra. In this work we will  focus on  $U(1)_k \times U(1)_{-k}$
Chern-Simons theory, which realizes the  chiral algebra of a rational compact boson (or `rational torus' in the terminology of \cite{Moore:1988ss}) as we will review in more detail in Section \ref{Secratboson}.

These Chern-Simons theories contain  in some sense a gravitational sector, since they all give rise to a boundary stress tensor generating left- and right- moving Virasoro subalgebras.
It is then natural  
to attempt to make  gravity fully dynamical by supplementing the bulk theory with a prescription to sum over topologies in the path integral. %Which geometries to include is however not a priori clear, since we are generically far  from the semiclassical, large $c$ regime. 
The ad-hoc prescription which is most commonly used in the literature, and which we will adopt also here,  includes, for a CFT ensemble defined on a Riemann surface $\S$, a sum over handlebody topologies whose boundary is $\S$.

For example, when $\S$ is a torus with complex structure parameter $\t$, this prescription tells us to sum over the $SL(2,\ZZ)$ family \cite{Maldacena:1998bw} of  handlebodies corresponding to the inequivalent ways of `filling in' the boundary torus. 
This leads to the bulk expression for the ensemble-averaged 
partition function (cfr. (\ref{MWaverage})) as a Poincar\'e sum 
\be 
\langle Z[\t] \rangle = \sum_{\g \in \G/\G_f} Z^{(1)}_{\rm seed} [\g \t] .\label{Zexpbulk}
\ee
Here, the 1-boundary seed amplitude arises from a Chern-Simons path integral on a specific solid torus, and  $\G_f$ is the subgroup of the modular group $\G$ which leaves $Z^{(1)}_{\rm seed} [\t]$ invariant.  We note that this expression is sensible, as the RHS is modular invariant by construction, and expanding it in a basis (\ref{ZI}) of modular invariants determines the ensemble weights $\r_I$.  

Note that, once the weights $\r_I$ are determined, so are all the connected $n$-point functions (\ref{Znconn}) in the ensemble. On the bulk side, the latter should come from path integrals on geometries connecting $n$ boundary tori.
The above prescription expresses them also in terms of  Poincar\'e sums of seed amplitudes of the form
\be
Z_{\rm conn}^{(n)}[\t_1, \ldots , \t_n]  = \sum_{(\g_1, \ldots, \g_n) \in \G^n/\G_f} Z^{(n)}_{\rm seed} [\g_1\t_1, \ldots , \g_n\t_n], \qquad n\geq 2.\label{Znconnbulk}
\ee
  Here, the seed amplitude $Z^{(n)}_{\rm seed}$ arises from a Chern-Simons path integral on a specific 3D geometry connecting 
  the $n$ boundary tori, and the Poincar\'e sums arise from performing relative Dehn twists between the boundaries. As we shall illustrate below, the  seed amplitudes satisfy certain geometric constraints,  and the RHS of (\ref{Znconnbulk}) is not  arbitrarily adjustable.
  Therefore,  nontrivial consistency conditions arise from equating (\ref{Znconnbulk}) with the boundary  ensemble expression (\ref{Znconn}). %, ultimately restricting the allowed  ensemble weights $\r_I$.
 
 In this work we will focus on  the constraints imposed by the ensemble interpretation of the two-boundary ($n=2$) amplitude. Besides strongly constraining the allowed 1-boundary seed amplitude $ Z^{(1)}_{\rm seed}$  and the ensemble weights, the solution will also instruct us to include contributions  from topological surface defects in the  2-boundary seed  $ Z^{(n)}_{\rm seed}$.

\subsection{Ensemble weights }
Let us make the above general statements more concrete, starting from the calculation of the ensemble weights from the relation (\ref{Zexpbulk}), following \cite{Castro:2011zq,Meruliya:2021utr}. The seed amplitude $ Z^{(1)}_{\rm seed}$ arises from a Chern-Simons path integral on a solid torus.
{The most natural choice is to perform this path integral   on the pure solid torus,   without any  Wilson line insertions.}
Viewing this geometry as a disk times a circle representing Euclidean time, one can obtain \(Z_{\mathrm{seed}}^{(1)}\propto|Z_{D_{2}\times S^{1}}^{CS}|^{2}\) from canonical quantization of two copies of \(U(1)_{k}\) Chern-Simons theory on the disk. Each copy leads to a chiral algebra \(\mathcal{A}\) on the boundary and a Hilbert space given by its vacuum irreducible representation {$\calr_0$}  \cite{Elitzur:1989nr}. Consequently the seed amplitude is proportional to the square of the vacuum character
\be
Z^{(1)}_{\mathrm{seed}} =\tilde{n}|\chi_0 |^2, \label{1bdyseedvac}
\ee
where $\tilde{n}$ is an as yet undetermined normalization. A possible generalization \cite{Meruliya:2021utr} of \eqref{1bdyseedvac} is to take as a starting point a more general combination of characters determined by a seed matrix $M_{\rm seed}$: 
\be 
Z^{(1)}_{\rm seed}  = \tilde  n \chi(\t)^\dagger M_{\rm seed} \chi(\t )  .\label{1bdyseedgen}
\ee
Since the quantization of the \(U(1)_k\) theory on the disk pierced by a Wilson line of charge \(j\in \mathbb{Z}_{2k}\) produces the representation $\calr_j$, the seed amplitude (\ref{1bdyseedgen}) can be realized as a
path integral on a solid torus with a  {formal} combination of Wilson  lines inserted along the noncontractible circle. % Quantization of the \(U(1)_k\) theory in presence of a line of charge \(j\in \mathbb{Z}_{2k}\) produces as a Hilbert space the \(j\)-th irreducible representation of \(\mathcal{A}\) and as a consequence a generic integer combination of lines for the two factors would lead to a seed amplitude of the type \comment{General seed matrix corresponds rather to a formal superposition of lines.}
However, we will see that not every seed matrix will lead to sensible ensemble weights satisfying (\ref{rhonorm}).
Note that the vacuum seed (\ref{1bdyseedvac}) is the special case
where 
\be 
M_{\rm seed}^{ij} = \d^i_0 \d^j_0 . \label{vacseedmatr}
\ee

Given a seed amplitude, the corresponding ensemble weights $\r_I$ are determined by combining eqs. (\ref{ensembleweights},\ref{Zexpbulk}). The result  can be written  as a matrix equation
\be 
\tilde n \sum_{\g \in  \G / \G_f } U_\g^{-1} M_{\rm seed}  U_\g =\sum_I \r_I M_I,
\ee
 Multiplying by $M_J$, taking the trace and using modular invariance of the $M_J$ matrices one finds
\be 
\r_I =  n  \sum_J (D^{-1})_{IJ} V^{\rm seed}_J, \label{rhosgen}
\ee
where $n = |\G/\G_f| \tilde n$ and we have defined the `seed vector' 
\be 
 V^{\rm seed}_I = \tr M_{\rm seed} M_I
 \ee
and the $\calm \times \calm$ matrix
\be 
D_{IJ} = \tr M_I M_J.\label{Ddef}
\ee
The normalization factor $n$ should ensure that (\ref{rhonorm}) holds, in particular
\be  n = \left( \sum_{K,L} (D^{-1})_{KL} V^{\rm seed}_L \right)^{-1} . \ee
One obvious constraint on the allowed seed matrices (\ref{1bdyseedgen}) is   that this expression is finite.
In the theory based on the vacuum seed (\ref{vacseedmatr}) we have, due to (\ref{vacunique}),
\be 
V^{\rm seed}_I = (M_I)^{00} = 1, \qquad \forall I.\label{vacseedvector}
\ee
%\subsection{Two-boundary  partition function}

\subsection{Two-boundary consistency conditions}
Having discussed the bulk partition function in the presence a single torus boundary and the determination of the ensemble weights,  let us now move on to the situation with two toroidal boundaries and % path integral on a manifold connecting two toroidal boundaries.
 study the implications of the equality of the bulk (\ref{Znconnbulk}) and boundary  (\ref{Znconn}) expressions for  this connected 2-point amplitude. In (\ref{Znconnbulk}) with $n=2$,  the 2-boundary seed amplitude \(Z_{\mathrm{seed}}^{(2)}\) should arise from a $U(1)_k \times U(1)_{-k}$ Chern-Simons path integral on a 3-geometry  with the topology of a torus times an interval. Therefore, we will assume  that it takes a holomorphically factorized form, i.e. 
 \be 
Z_{\mathrm{seed}}^{(2)}  \propto |Z_{T^2\times I}|^2. \label{holfact}
 \ee
Furthermore,  it was argued on general grounds in \cite{Cotler:2020ugk} (see also \cite{Collier:2021rsn}) that $Z_{T^2\times I} (\t_1, \t_2)$  should be invariant under a simultaneous modular transformation of the arguments in the   sense that
\be 
Z_{T^2\times I} (\g \t_1,\S \g \S  \t_2) = Z_{T^2\times I} (\t_1, \t_2),\label{whmodprop}
\ee
where 
\be 
\S = \left( \begin{array}{cc} -1 &0\\0 & 1\end{array}\right)
\ee
implements a sign reversal of $\t$. 
The argument goes as follows (see Figure \ref{Fig:cotler-jensen-argument}):
\begin{figure}
\centering
   \begin{subfigure}[t]{.4\textwidth}
     \centering
     \includegraphics[width=\textwidth]{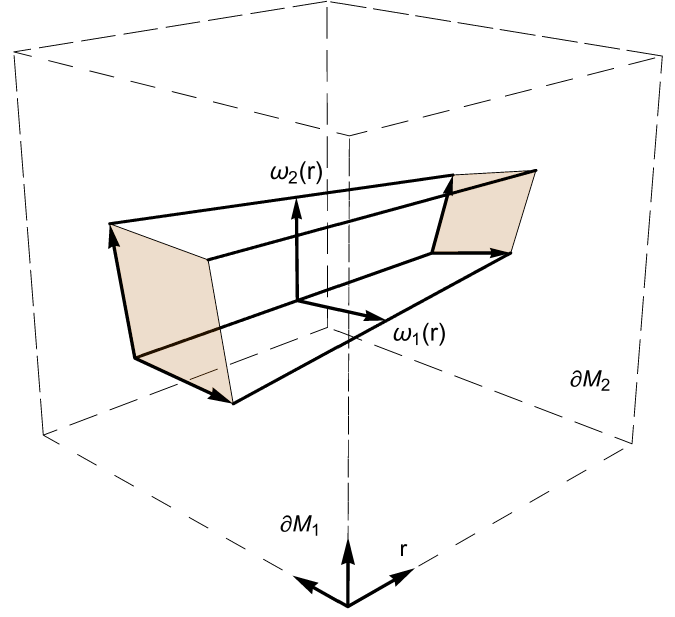}
     \caption{}
   \end{subfigure}
   \qquad
   \begin{subfigure}[t]{.4\textwidth}
     \centering
     \includegraphics[width=\textwidth]{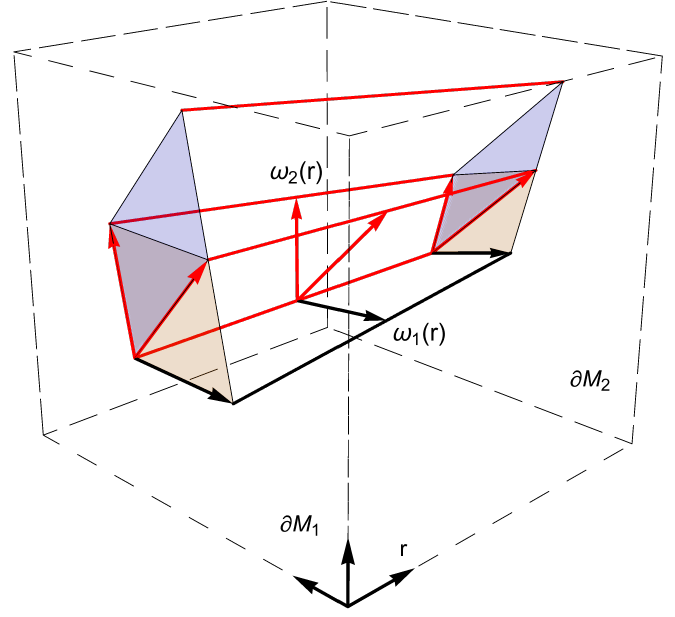}
     \caption{}
   \end{subfigure}
\caption{{ (a) The two boundaries of \(M=T^2\times I\) have complex structures induced by choices of lattice bases \(\omega_{1,2}(r=0)\) and \(\omega_{1,2}(r=1)\) (\(r\) is the coordinate on the \lq radial\rq\ direction \(I\)). (b) A simultaneous modular transformation on the two boundaries is induced by the same change of basis on the two lattices and can be smoothly transported as a large diffeomorphism through the bulk (the transformed bases are colored in red).}}
\label{Fig:cotler-jensen-argument}
\end{figure}
a modular transformation on one  boundary torus stems from a different choice of basis for the lattice defining it, and this new basis is transported smoothly through the bulk to induce  the same change of basis in the second boundary. Therefore the Chern-Simons path-integral  $\tilde Z_{T^2 \times I} (\t_1 , \bar \t_2) = \tr q_1^{L_0^{(1)}}  \bar q_2^{L_0^{(2)}}$ should be invariant under simultaneous modular transformations 
\be 
\tilde Z_{T^2 \times I}  ( \g \t_1 ,\g  \bar \t_2) =  \tilde Z_{T^2 \times I}  (  \t_1 ,  \bar \t_2) .\label{CJsameor}
\ee
In this path integral, the two  boundaries are oppositely oriented as induced from the bulk. To obtain the partition function for boundary  tori with the   same orientation, we replace $\t_2 \to - \bar \t_2$, leading to 
\be
Z_{T^2 \times I} ( \t_1,\t_2)  =  \tilde Z_{T^2 \times I}  ( \t_1 ,- \t_2).
\ee
It is straightforward to check that the property (\ref{CJsameor}) translates into (\ref{whmodprop}). 
 
% The conjugation by $\S$ in the first argument is due to the fact that we take both boundaries to have the same orientation, while the orientations induced from the bulk are opposite. 
The property (\ref{whmodprop}) implies that we can expand the seed amplitude   in terms of  $\calm$ unknown `wormhole  coefficients' $c_I$ as follows: %multiplying modular invariant matrices { 
\be 
Z_{T^2\times I} ( \t_1, \t_2) = \sum_I c_I \chi (\t_1)^T M_I  \chi (\t_2).\label{whexp}
\ee
The equality of the expressions  (\ref{Znconn}) and  (\ref{Znconnbulk})    can once again be expressed as a matrix identity, which can be reduced to
\be 
%\r_K \d_{KL} - \r_K \r_L = | \G : \G_f| (D^{-1})_{KI}  (D^{-1})_{LJ}  \bar c_I  c_J  D_{IKJL},\label{cIJ}
\r_I \d_{IJ} - \r_I \r_J = \sum_{M,N,K,L} \bar c_M  c_N  (D^{-1})_{IK}  (D^{-1})_{JL}  D_{MKNL},\label{mastereq}
\ee
where we defined
\be 
 D_{IJKL} = \tr M_I M_J M_K M_L.
\ee
To derive (\ref{mastereq}) we have used our assumption (see below (\ref{vacunique}))  that the $M_I$ are real and symmetric matrices. 

We should note that  (\ref{mastereq}) is in general an overdetermined system %of $\calm^2$ equations.\footnote{ We note that symmetry of the LHS impose that $c_M \bar c_N$ is real for all $M,N$.} 
for the $\calm$ complex coefficients $c_I$,  and it is therefore to be  expected that the existence of solutions will also impose constraints on the $\r_I$ and therefore the 1-boundary seed amplitude $Z^{(1)}_{\rm seed}$. 
% In the more general case I get that we should define 
%\be 
%D_{IJKL} = \tr M_I^\dagger M_J^t M_K M_L.
%\ee}
\subsection{Wormhole amplitude from canonical quantization}\label{Sec:wormhole_from_canonical_qzn} 
Naively one might expect that the 2-boundary seed amplitude $Z^{(2)}_{\rm seed}$ should be given by the {  pure} Chern-Simons path integral on the $T^2 \times I$ wormhole geometry. This path integral is the RCFT equivalent of the wormhole partition function computed by Cotler and Jensen for  pure gravity and noncompact $U(1)$ Chern-Simons theories \cite{Cotler:2020ugk,Cotler:2020hgz}. However, as we shall see, this Chern-Simons path integral, which we will denote as
$Z_{T^2 \times I}^{\rm CS}(\t_1, \t_2)$, gives only part of the required answer and needs to be supplemented with additional `exotic' wormhole contributions which it will be our task to identify.

Let us first recall the result for the  pure Chern-Simons path integral  $Z_{T^2 \times I}^{\rm CS}(\t_1, \t_2)$ from the standard Chern-Simons/RCFT correspondence. Viewing  $T^2 \times I$ as an annulus $S^1 \times I$ times a circle representing Euclidean time, we can obtain $Z_{T^2 \times I}^{\rm CS}(\t_1, \t_2)$ from the canonical quantization of the theory on the annulus. In this case there are chiral algebras $\cala$ and $\overline \cala$ associated to   the two boundary circles of the annulus,  and \cite{Elitzur:1989nr} obtained the following result for the  Hilbert space:
\be 
\calh_{ S^1 \times I } = \sum_{i = 0}^{\caln-1}  \calr_i \otimes \overline {\calr_i}, \label{Hilb-sp_annulus}
\ee
where we recall that  the $\calr_i$ are the irreducible representations of the chiral algebra. Correspondingly, the path intregral    is given by (allowing for an undetermined normalization factor $n$)
\be 
 Z_{T^2 \times I}^{\rm CS}(\t_1, \t_2) = n  \sum_{i = 0}^{\caln-1} \chi_i (\t_1) \chi_i (\t_2). \label{pureCS_PI_annulus}
 \ee
Here, we have followed the steps below (\ref{CJsameor}) to account for boundaries with the same orientation.  
 As a consistency check, we see that this is indeed of the general form (\ref{whexp}): it is the particular case  where only the diagonal modular invariant $M_1^{ij} = \d^{ij}$ appears. In other words, if the pure Chern-Simons partition function were the full answer, the coefficients $c_I$ would be of the form
 \be 
 c_I^{\rm CS} = n  \d_I^1.\label{cpureCS}
 \ee

 One remark is in order before we continue to discuss a particular class of RCFT ensembles. For simplicity, we have in this section considered CFT partition functions and characters which only keep track of  the Virasoro conformal weights $L_0$ and $\bar L_0$. However, when the chiral algebra is extended, it can be desirable to work with refined partition functions and characters which keep track of further quantum numbers (commuting with $L_0, \bar L_0$) and depend on additional chemical potentials.  The formulas in this section readily generalize to this setting, provided we assign the proper modular transformation law to the chemical potentials, as  we will illustrate in the examples below.

\section{Rational boson ensembles and their bulk duals}\label{Secratboson}
In what follows we will make the  general considerations of the previous Section explicit in a simple 
 class of ensemble holographic theories. In these, the exotic gravity theory is based %These are the exotic gravity models  based  
 on a compact $U(1)_k\times U(1)_{-k}$ Chern-Simons theory with
 action
\be 
S[A, \tilde A] = - {k \over 2\p} \int_\calm ( A\wedge dA - \tilde A \wedge d \tilde A),\label{SCS}
\ee
where the level\footnote{Our normalization of the level is chosen in order to avoid factors of 2 in formulas below. Many references use a differently  normalized level  related to ours as $k_{\rm there} = 2k$.} $k$ should be an integer in order  for the path integral to be well-defined. 
 As shown in \cite{Raeymaekers:2021ypf}, a large subset of these remains under analytic control even  when the size of the ensemble goes to infinity. This will allow us to find explicit expressions for the ensemble weights (\ref{rhosgen}) and to investigate the 2-boundary consistency conditions (\ref{mastereq}).  
 
 \subsection{Chiral algebra and representations}
 The  chiral algebra of the models of interest, which we denote  as $\cala_k$, arises in free boson CFTs at rational values of the radius squared. 
Indeed, let us consider a compact free boson at radius\footnote{Our compact boson conventions follow Polchinski's book \cite{Polchinski:1998rq} with $\a '=1$.} $R^2 = p/q$ with $p$ and $q$ relatively prime and positive integers. Setting $k = pq$ this theory has chiral currents
\be 
J = \pa X (z) , \qquad W_\pm = : e^{\pm 2 i \sqrt{k} X(z)} :
\ee 
with conformal weights 1 and $k$ respectively. These generate the chiral algebra $\cala_k$.

This algebra has { $2k$ } inequivalent { irreducible} representations $\calr_j, j =0 , \ldots , 2k-1$ \cite{Moore:1988ss} (for a review, see  \cite{DiFrancesco:1997nk}). 
To fully  characterize these, we should not only keep track of their weights under the Virasoro generator $L_0$ but also under the $u(1)$ generator $J_0$. That is, we will work with   refined characters defined as
\bea
\chi_j (\t, z) &:=& \tr q^{L_0} \z^{J_0}\\
&=&{1 \over \h (\t)}  \sum_{n \in \ZZ} q^{(2 nk + j)^2\over 4k }\z^{ 2nk + j \over 2}
\eea
where $q = e^{2 \p i \t}, \z = e^{-2\p i z}$, and
$\h = q^{1\over 24}\prod_{m=1}^\infty (1- q^m)$ is Dedekind's eta function. The characters satisfy 
\be 
\chi_{ 2k + j} = \chi_j
\ee
so we can take $j$ in the range $j = 0, \ldots , 2k-1$. Charge conjugation sends $j \to - j$:
\be 
\chi_{j*} (\t,z) = \chi_{-j}  (\t,z).
\ee
We note that the representations $\calr_0$ and $\calr_k$ are self-conjugate, while the remaining $\calr_j$ with  $j \notin  k \ZZ$ are not, since $\chi_{j*} (\t,z) \neq \chi_{j} (\t,z)$. However, the specialized characters evaluated  at $z=0$ cannot distinguish conjugate representations:
\be 
\chi_{j*} (\t,0) = \chi_{j} (\t,0).\label{conjprop}
\ee
Therefore, in order to keep track of the full  $\cala_k$ representation content we need  to work with refined characters rather than specialized ones.

Under modular transformations $\g = \tiny{ \left( \begin{array}{cc} a &b \\ c& d\end{array}\right)} \in SL(2,\ZZ)$, the characters transform as  
\begin{equation} 
\chi  \left( {a \t + b \over c \t  + d},{z\over c\t + d}\right) = 
e^{\frac{i \pi k}{2} \frac{c z^2}{c\t + d}} U_\g \cdot \chi( \t, z).
\end{equation}
Here, the  $U_\g$ furnish a $2k$-dimensional unitary representation of $SL(2, \ZZ)$. In particular, the modular $S$ and $T$ transformations are represented as  
\begin{align}
T:  && \chi_{j} (\t+1 , z) =& e^{ 2 i \p \left({j^2 \over 4 k} - {1 \over 24}\right)}  \chi_{j} (\t  , z) \equiv (\calt \chi)_j , \\
	S:  && \chi_{j} \left(- {1 \over \t} , {z \over \t}\right) =& e^{\frac{i \pi k z^2}{2 \t}} {1 \over \sqrt{2 k}} \sum_{j'= 0 }^{2k-1} e^{- 2 i \p {j j'\over 2k}} \chi_{j', k} (\t , z) \equiv (\cals \chi)_j.\label{modularS}
\end{align}
One checks that the following group relations hold:
\begin{equation} 
(\cals \calt)^3 = \cals^2 = \calc , \qquad \calc^2 = 1.
\end{equation}
Since  the charge conjugation matrix \(\calc\) is not the identity, the matrices $U_\g$ furnish a representation of $SL(2, \ZZ)$ rather than $PSL(2, \ZZ)$.

Due to the $z$-dependent prefactor in (\ref{modularS}), the refined characters do not transform in a matrix representation of $SL(2,\ZZ)$ under modular transformations. This can be remedied by multiplying the characters by an extra factor whose transformation offsets the anomalous part. Indeed, one checks that 
\be 
\tilde \chi_j (\t,z) = e^{  \p k{ z^2\over 4 \t_2}}\chi_j (\t,z)
\ee
transforms in the unitary matrix representation of $SL(2,\ZZ)$ furnished by the matrices $U_\g$.

We should also note that the   rescaled   characters $\tilde \chi (\t, z)$ are the ones that naturally  appear in  expressions obtained from a  covariant path integral formulation \cite{Kraus:2006nb}. In the 2D free boson theory, the chemical potential $z$ is introduced in the Lagrangian by adding a term of the form $\cala_{\bar w}\pa_w X + \cala_w \pa_{\bar w} X$, where $\cala_{\bar w} \propto z $ . In converting to  Hamiltonian form one gets a cross term proportional to $\cala_w \cala_{\bar w}$ which is precisely the prefactor in (\ref{ZRref}). The argument extends   to  3D Chern-Simons path integrals, e.g. the path integral on the solid torus with appropriate boundary conditions is proportional to $|\tilde \chi_0 (\t ,z) |^2$ \cite{Datta:2021ftn}.

\subsection{Modular invariants}
Let us now discuss the modular invariant combinations of  $\cala_k \times \overline{\cala}_k$ characters.
%To describe the modular invariants  of the theory, we note
 From the construction above we note that, for each divisor $\d$ of $k$ (which we will write as $\d | k$ in what follows), the compact boson at radius $R^2 = k/\d^2$ yields a realization of the algebra  $\cala_k \times \overline{\cala}_k$. The refined and  rescaled
 	 partition function\footnote{The notation  $[ \t, z]$ is shorthand for $( \t, \bar \t, z, \bar z)$.}
 \be 
 Z_R [ \t, z] = e^{ \p k {z^2+ \bar z^2\over 4 \t_2}} \tr_{\calh_R} 
 q^{L_0} \bar q^{\tilde L_0} \z^{J_0} \bar \z^{\tilde J_0}\label{ZRref}
 \ee
 therefore furnishes a modular invariant. Furthermore, a full analysis of the commutant of the $\cals$- and $\calt$-matrices \cite{Cappelli:1986hf,Cappelli:1987xt} shows  that the modular invariants obtained in this way form a complete basis. It follows that in these theories, the number of independent  modular invariants is $\calm = d (k)$, the number of distinct divisors of $k$.  
 
 It will be convenient to introduce a specific labelling of the modular invariant partition functions.  For this  we first label the divisors of $k$ with an index $I$ running from 1 to $d(k)$ in increasing order %such that they increase in magnitude
 \be 
 1 = \d_1 < \d_2 \ldots < \d_{d(k)}.\label{deltalabels}
 \ee
 We will use the same label to indicate the  modular invariant  partition function associated to $\d_I$, i.e.
 \be 
 Z_I  [\t,z ] \equiv   Z_{R={\sqrt{k} \over \d_I} }[\t,z ].
 \ee
As before we associate to $Z_I$ a modular invariant $2k \times 2k$ matrix $M_I$ through decomposition in (rescaled) characters
 \be 
Z_I  [\t,z ] =  \tilde \chi (\t,z)^\dagger M_I \tilde \chi (\t,z).\label{Mmatrixref}
\ee
Let us now give an explicit expression for these matrices $M_I$.
 
 It will be useful to introduce the symbol $\a_I$ to denote the  greatest common denominator of $\d_I$ and $k/\d_I$:
 \be
 \a_I :=\gcd  \left(\d_I, {k\over \d_I} \right).\label{alphadef}
 \ee
 Note that $\a_I$ is a  quadratic  divisor of $k$, $\a_I^2 |k$.
%We also define a quantity $\o_I$ as follows.
 The definition (\ref{alphadef}) and Bezout's lemma imply that there exist integers $r_I,s_I$ such that \be r_I  {\d_I \over \a_I} - s_I  {k \over \a_I \d_I} = 1.\label{rsdef}\ee 
 From these we define the quantity $\o_I$ as follows
 \be 
 \o_I = \left[ r_I  {\d_I \over \a_I} + s_I  {k \over \a_I \d_I}\right]_{2k\over \a_I^2} ,
 \label{omdef}
 \ee
 where we introduced the notation $ [x]_n := x \mod n$.
We note that  $\o_I$ is defined modulo $2 k/\a_I^2$, and it is straightforward to  check that a different choice of $r_I$ and $s_I$ leads to the same  $\o_I$. Furthermore, $\o_I$ are roots of unity modulo $4 k / \a_I^2$,
\be \o_I^2 := 1 \mod {{4 k \over \a_I^2} }\label{osqprop}.
\ee
With these definitions, the components of the modular invariant matrices $M_I$ in (\ref{Mmatrixref}) are \cite{Raeymaekers:2021ypf}  
 \be 
 M_I^{ij} =\left\{ \begin{array}{ll} {1\over \a_I} \d_{[i- \o_I j]_{2 k\over \a_I}}&  {\ \rm if \ } \a_I| i  { \rm \ and\ }  \a_I | j  \\
 	0 &{\ \rm otherwise}  \end{array} \right., \qquad 0 \leq i, j< 2k . \label{Mcomp}
 \ee
 Here, the quantity $\d_{[x]_n}$ is defined to be one if $x\equiv 0\ (\md n)$ and zero otherwise. We note that all these modular invariants are physical in the sense of Section \ref{SecRCFTens}: the coefficients in the character decomposition are positive integers and    the vacuum representation occurs precisely once, $M_I^{00} =1, \forall I$.
We note that the uniqueness of the vacuum fixes the normalization of the modular matrices, preventing us from multiplying them by some positive natural number.
 
 We close this subsection with a remark on T-duality. Since, in our labelling (\ref{deltalabels}), $\o_1 = 1$ and $\o_{d(k)}=-1$,  the corresponding  modular invariant matrices are 
 \be
 M_1 = 1, \qquad M_{d(k)}  = \calc.
 \ee
 These cases correspond to the diagonal and to the charge conjugation  modular invariant respectively. More generally, 
  the root of unity associated to $k/ \d_I$ is   $-\o_I$ and    their modular matrices are related as
 \be 
 M_{\d_I} = \calc M_{k\over \d_I}.
 \ee
 Therefore, due to  (\ref{conjprop}), at vanishing chemical potential their partition functions are the same,
 \be 
  Z_{\sqrt{k} \over \d_I  }[\t, 0] =  Z_{ \d_I \over \sqrt{k}  }[\t, 0]
 \ee
 which expresses the standard T-duality invariance of the compact boson partition function.

\subsection{Monoidal multiplicative structure}
Before proceeding, let us comment on the further structure on the  space of the modular invariant matrices.  From the component form (\ref{Mcomp}) one can deduce that multiplying two modular invariant matrices gives again a modular invariant matrix. Concretely (see \cite{Raeymaekers:2021ypf} for details) one finds
\be 
M (\o_I ) M(\o_J) = \gcd(\a_I, \a_J ) M \left(   \o_I \bullet \o_J  \right)\label{Mmultgen}
\ee
where $\bullet$ stands for  the following multiplication rule 
\be
 \o_I \bullet \o_J = [ \o_I \o_J ]_{2 k\over {\rm lcm} (\a_I, \a_J)^2}.\label{ommult}
	\ee
The prefactor on the RHS could, at the cost of working with non-physical modular matrices,  be absorbed in a rescaling of the $M_I$  by a factor $\a_I^{-1}$; one then sees that  the  rescaled matrices %\footnote{Recall that the normalization of our modular matrices was fixed by requiring that they represent physical modular invariants, leading to the multiplication law (\ref{Mmultgen}).}  
form a monoid under multiplication.  This is consistent with the fact that modular invariants are in one-to-one correspondence with topological surface operators in the Chern-Simons theory. As we shall show  in detail in Section \ref{Secdefects}, the multiplication rule (\ref{Mmultgen}) agrees with the fusion rule for these defects, which define a monoidal two-category.

This additional structure becomes even more restrictive in the case where $k$ is square-free, meaning that it has no non-trivial square divisors $\b^2 | k, \b>1$, so that the numbers $\a_I$ defined above are  all equal to one. In this case the monoidal structure simplifies to  an actual  group structure. More precisely, the modular matrices form a  representation of  the multiplicative group $G_k$ of roots of unity modulo $2k$, 
\be 
G_k =   \{  [\o ]_{ 2k};   \o^2 \equiv 1\  (\md 2 k) \}.\label{Gdef}
\ee
To show this, we first note that, for the modular invariant associated to the divisor $\d_I$, the 
  quantity \(\omega_I\) defined in \eqref{omdef} with    \(\alpha_I = 1\)  is   an element of \(G_k\)  {due to (\ref{osqprop}).} 
 Conversely, to every element \(\omega\in G_k\) one can associate the divisor \be 
\delta(\omega) := \gcd\left(\frac{\omega+1}{2} , k\right).
\label{omdef-inv-sqfree}
\ee
Notice that, by definition, \(\omega\) is its own multiplicative inverse modulo \(2k\) hence it must be coprime with \(2k\). So every \(\omega\in G_k \) is in particular odd and the quantity \(\frac{\omega+1}{2}\) is an integer. Moreover \(\delta(\omega)\) is well defined even if \(\omega\) is defined modulo \(2k\), because of the property \(\gcd(a+mk,k)=\gcd(a,k)\). One can check that the definitions \eqref{omdef} and \eqref{omdef-inv-sqfree} are each others inverse, providing a one-to-one correspondence between the set of divisors of \(k\) and the elements of \(G_k\). {  In this special case, from the component expression (\ref{Mcomp}) we see that the modular invariant matrices $M_I$ furnish the  $2k$-dimensional representation  corresponding simply to the  multiplication by $\o_I$ in $\ZZ_{2k}$.

For generic \(k\), { the multiplication rules (\ref{Mmultgen}) show that} the  modular matrices form a representation of  {a commutative  monoid with unit element. It can  be seen to possess the following structure}\footnote{ Note that in general we cannot replace \(\mathrm{mod} 4k/\alpha^2\) with \(\mathrm{mod} 2k/\alpha^2\) in \eqref{omdef-inv} when \(k\) is not square-free. This can be seen noticing that the group of square roots of unity modulo \(2k/\alpha^2\) has \(2^{\Omega(k/\alpha^2)}\) elements when \(k/\alpha^2\) is odd or an odd multiple of 2, otherwise it has \(2^{\Omega(k/\alpha^2)+1}\) elements \cite{Omami:2009}. Instead notice that \(G^\alpha_k\) is generically only a \(\mathbb{Z}_2\) quotient of the group of square roots of unity modulo \(4k/\alpha^2\) (if \(\omega\) is a root of unity modulo \(4k/\alpha^2\), also \(\omega+2k\) is a distinct root). The latter has \(2^{\Omega(k/\alpha^2)+1}\) elements for any value of \(k/\alpha^2\), so \(|G_k^\alpha|=2^{\Omega(k/\alpha^2)}\) which is generically different from the number of roots of unity modulo \(2k/\alpha^2\). For square-free \(k\) (so \(\alpha=1\)) instead the two cardinalities always agree, justifying \eqref{omdef-inv-sqfree}.}%\footnote{{\color{blue} Note that in general we cannot replace \(\mathrm{mod} 4k/\alpha^2\) with \(\mathrm{mod} 2k/\alpha^2\) in \eqref{omdef-inv}, contrarily to \eqref{omdef-inv-sqfree}, when \(k\) is not square-free. This is because if \(\omega^2 = 1\mod 2q\) for some number \(q\), then there is an integer \(N\) such that \((\omega-1)(\omega+1)=2qN\). Since \(\omega\) must be odd, \((\omega\pm 1)\) are two consecutive even number hence one of the two is divisible by 4. By integrality of the above equation, this forces the combination \(qN\) to be divisible by 4. If \(q=k\) is square-free, it is at most divisible by 2, so \(N\) must be even and we have \(\omega^2=1\mod 4k\tilde{N}\) in agreement with \eqref{omdef-inv-sqfree}. If instead \(q=k/\alpha^2\) is any integer, \(N\) can very well be odd, hence we cannot replace \(2q\) with \(4q\) in \eqref{omdef-inv}.}}
\begin{equation}
G_k = \bigcup_{\alpha\in \mathrm{SqDiv}(k)} G^\alpha_k ,\qquad G_k^\alpha := \{  [\o ]_{ 2k/\alpha^2};   \o^2 \equiv 1\  (\md 4 k/\alpha^2) \}.
\label{omdef-inv}
\end{equation}
Here, corresponding to each square-divisor \(\alpha\) there is a maximal subgroup \(G^\alpha_k\) which is one-to-one with the subset of divisors of \(k\) satisfying \(\gcd(\delta,k/\delta)=\alpha\). This can be shown similarly to above, generalizing \eqref{omdef-inv-sqfree} to 
\be
\omega\in G^\alpha_k :\qquad \delta(\omega) := \alpha \gcd\left(\frac{\omega+1}{2} , \frac{k}{\alpha^2}\right)
\ee
which inverts \eqref{omdef} in the general case. This provides a one-to-one correspondence between the set of divisors of \(k\) and \(G_k\) in full generality.
}

 \section{Analysis of the two-boundary constraints}\label{Secsolconstr}
 Having set the stage, we will in the rest of this work explore the  solutions to the equations (\ref{rhosgen}) determining the ensemble weights and the two-boundary  consistency conditions (\ref{mastereq}). % on the 2-boundary seed coefficients $c_I$.  
\subsection{Examples for small $k$}
 It will prove instructive to first consider the problem for a few simple cases at low values of $k$ as we shall do presently; in the next  subsection we will then find the general solution  for square-free $k$.
\subsubsection{$k=1$}
In this simplest example, there are two irreducible representations $(\caln =2$) and only one possible modular invariant $(\calm =1)$, the diagonal one. Hence there is no ensemble to average over and (\ref{mastereq}) indeed tells us that the wormhole coefficient $c_1$ should vanish. 
\subsubsection{$k=2$}
In this example there are four   representations $(\caln =4$) and two modular invariants $(\calm =2$). In our labelling convention we have 
\be \d_I = (1,2), \qquad \a_I = (1,1), \qquad \o_I = \left([1]_{4},[-1]_4\right),\ee and the modular invariant matrices are  the identity and the charge conjugation matrix $\calc$ 
\be 
M_1 = 1, \qquad M_2 =
\calc  = {\small\left( \begin{array}{cccc} 
	1&0&0&0 \\
	0&0&0&1\\
	0&0&1&0 \\
	0&1&0&0 
\end{array}\right)}.
\ee
Since $\calc^2=1$ the group they form under multiplication is isomorphic to $\ZZ_2$. The matrix $D_{IJ}$ defined in (\ref{Ddef}) is equal to 
\be
D_{IJ} = {\small \left( \begin{array}{cc} 4 & 2\\2 &4
\end{array}\right)} .
\ee
Starting from a general seed vector
\be 
V^{\rm seed}_I =(v_1, v_2)
\ee
 we find  from (\ref{rhosgen}) the ensemble weights 
\be 
%\r_I = (V^{\rm seed}_1 + V^{\rm seed}_2)^{-1} ( 2 V^{\rm seed}_1 - V^{\rm seed}_2,  2 V^{\rm seed}_2 - V^{\rm seed}_1 ).
%\r_I = \left( {2 V^{\rm seed}_1 - V^{\rm seed}_2\over  V^{\rm seed}_1 + V^{\rm seed}_2},  {2 V^{\rm seed}_2 - V^{\rm seed}_1\over V^{\rm seed}_1 + V^{\rm seed}_2} \right).
\r_I = \left( {2v_1 - v_2\over  v_1 +v_2},  {2 v_2 - v_1\over v_1 + v_2} \right).\label{rhok2}
	\ee
In order for the weights	to take values between zero and one the seed vector needs to satisfy
\be
 \half \leq  {v_1 \over v_2} \leq 2.
 \ee
 The special case of  the vacuum seed vector $V^{\rm seed} = (1,1)$ (cfr.\ (\ref{vacseedvector}))  leads to equal ensemble weights, $\r_I = \half (1,1)$. 
 
 Next we turn to the equations (\ref{mastereq}) for the wormhole coefficients $c_I$. In this example the system is solvable for general ensemble weights, and the solution  is only determined up to an overall phase $\theta_2$:
 \be
 c_I = \sqrt{\r_1 \r_2} e^{i\theta_2} (1,-1).\label{cIk2}
 \ee
This example generalizes to the situation where $k$ is a prime number in a straightforward way.

The examples with only 1 or 2  modular invariants considered so far are special;  for a greater number of modular invariants  (\ref{mastereq}) also places constraints   on the ensemble weights $\r_I$,  as the two following examples show. 

\subsubsection{\bf $k=6$}
In this case, $\caln = 12, \calm = 4$ and we have
\be 
\d_I = (1,2,3,6), \qquad \a_I= 1, \qquad  \o_I= \left([ 1]_{12}, [7]_{12}, [5]_{12}, [11]_{12}\right).
\ee
Working out the modular invariant matrices from (\ref{Mcomp}) one finds that they form a group isomorphic to $(\ZZ_2)^2$, more precisely
\be 
M_1 \simeq (1,1),\qquad M_2 \simeq (-1,1), 
\qquad M_3 \simeq (1,-1) \qquad M_4 \simeq (-1,-1).\label{k6isom}
\ee
The matrix $D_{IJ}$ is found to be 
\be 
D_{IJ} = {\small \left(
\begin{array}{cccc}
	12 & 6 & 4 & 2 \\
	6 & 12 & 2 & 4 \\
	4 & 2 & 12 & 6 \\
	2 & 4 & 6 & 12 \\
\end{array}
\right)},
\ee
and a general 1-boundary seed vector $V^{\rm seed}_I =(v_1, v_2,v_3,v_4)$ leads to the ensemble weights
{\small\be 
\r_I=\left(\frac{6  {v_1}-3  {v_2}-2  {v_3}+ {v_4}}{2 ( {v_1}+ {v_2}+ {v_3}+ {v_4})},\frac{-3
	 {v_1}+6  {v_2}+ {v_3}-2  {v_4}}{2 ( {v_1}+ {v_2}+ {v_3}+ {v_4})},\frac{-2
	 {v_1}+ {v_2}+6  {v_3}-3  {v_4}}{2 ( {v_1}+ {v_2}+ {v_3}+ {v_4})},\frac{ {v_1}-2
	 {v_2}-3  {v_3}+6  {v_4}}{2 ( {v_1}+ {v_2}+ {v_3}+ {v_4})}\right). \label{rhok6}
\ee}
Positivity of the $\r_I$ again restricts the allowed ranges of the $v_I$.

Next we turn to the 2-boundary consistency constraints (\ref{mastereq}). 
We first convert to a basis in which the RHS of  (\ref{mastereq}) is diagonal. In order for solutions to exist,  the LHS must also be diagonal in this basis, which places restrictions on the allowed seed vector $V^{\rm seed}$. One finds that these lead  
to two  classes of admissible  seed vectors. Either it   is (up to an overall rescaling) of the form $v_I = \tr M_I M_J$, for some $J$. In this case, the  1-boundary seed amplitude is already modular    invariant and given by  $M_J$. This  picks out a single CFT theory, $\r_I = \d_{IJ}$, rather than an ensemble, and the wormhole coefficients $c_I$  consequently vanish. 

The second consistent possibility, and the only one leading to an ensemble of theories, is that  seed vector is (up to rescaling) $v_I = (1,1,1,1)$. This arises most naturally from the vacuum seed amplitude (\ref{vacseedmatr}), but it could also come from starting with the charge-$k$ excited state $Z^{(1)}_{\rm seed} = \tilde  n  |\tilde \chi_k |^2$.  We see from (\ref{rhok6})
 that the  seed vector $v_I = (1,1,1,1)$ leads to a maximal entropy ensemble with \be \r_I= \frac{1}{4}(1,1,1,1).\ee The solution for the corresponding
 wormhole coefficients is determined up to 3 phases, which for later comparison we label as $\theta_{2,3,4}$; one finds
 \bea 
c_I &=&\frac{1}{4} \left(   e^{i \theta_2}+\sqrt{2} e^{i \theta_3}+\sqrt{3} e^{i \theta_4} ,  
 -e^{i \theta_2}-\sqrt{2} e^{i \theta_3}+\sqrt{3} e^{i \theta_4} ,\right. \nonu
&& \left.   -e^{i \theta_2}+\sqrt{2} e^{i \theta_3}-\sqrt{3} e^{i \theta_4} ,    e^{i \theta_2}-\sqrt{2} e^{i
	\theta_3}-\sqrt{3} e^{i \theta_4} \right).\label{cIk6}
 \eea

\subsubsection{\bf $k=9$}\label{exk9}
In this example, $k$ is not square-free.  We have $\caln = 18, \calm = 3$ and
\be 
\d_I = (1,3,9),\qquad \a_I = (1,3,1), \qquad\o_I=  (1,1,17).
\ee 
The modular matrices form a   monoid under multiplication, with $M_1$ playing the role of an identity element
and
\be 
M_2 \cdot M_2 = 3 M_2,\qquad  M_2 \cdot M_3 = M_2, \qquad M_3 \cdot M_3  = M_1.
\ee
In other words, $M_2$ does not possess an inverse. Note that $M_1, M_3$ form a $\ZZ_2$ subgroup.

Denoting again $V^{\rm seed} = (v_1,v_2,v_3)$, the general ensemble weights are
\be 
\r_I =\left(\frac{3 (3  {v_1}- {v_2})}{2 (3  {v_1}+2  {v_2}+3  {v_3})},\frac{-3  {v_1}+10  {v_2}-3
	 {v_3}}{2 (3  {v_1}+2  {v_2}+3  {v_3})},-\frac{3 ( {v_2}-3  {v_3})}{2 (3  {v_1}+2
	 {v_2}+3  {v_3})}\right).\label{rhok9}
\ee
We note that, in contrast with the previous examples,  starting from the vacuum seed vector $v_I = (1,1,1)$ does not lead to a uniform distribution but to $\r_I = (3/8,1/4,3/8)$.

An analysis of the 2-boundary consistency conditions (\ref{mastereq})  for the wormhole coefficients shows that only two types of solutions are possible.  As in the previous example, a trivial class of solutions occurs when we start from a modular invariant 1-boundary seed amplitude, in which case there is no ensemble and the wormhole amplitude vanishes, $c_I =0$.
The second possibility occurs when the seed vector is such that the non-invertible invariant $M_2$ doesn't appear in the ensemble, i.e. $\r_2=0$. From (\ref{rhok9}) this happens for
\be 
v_2 = {3\over 10}(v_1 + v_3).
\ee
The analysis proceeds as in the $k=2$ case, as we are left  with an ensemble containing 2 modular invariants forming a $\ZZ_2$ group. The  solutions in this class have 
\be\begin{aligned}
\r_I &= \left(\frac{9  {v_1}- {v_3}}{8 ( {v_1}+ {v_3})},0,\frac{9  {v_3}- {v_1}}{8
	( {v_1}+ {v_3})}\right), \\
c_I &= \sqrt{ 8 \r_1 \r_3} e^{i \theta} (1,0,-1) .
\end{aligned}\ee
One can check that these exhaust  all consistent possibilities. We note in particular that starting from the vacuum seed $V^{\rm seed} \propto (1,1,1)$
is in this case ruled out. 

%and the general solution to (\ref{}) is

\subsection{General solution for square-free $k$}
The first three  examples above belong to the class where $k$ is square-free. We will now derive an explicit formula  the ensemble weights (\ref{rhosgen}) and analyze the consistent solutions to the 2-boundary constraints (\ref{mastereq}) within this class of rational boson ensembles.
%We now focus on the subclass of rational boson ensembles for which $k$ is square free.  of We will review the result of \cite{Raeymaekers:2021ypf} that the ensemble weights are in this case all equal, and will derive the general solution for the wormhole coefficients satisfying the consistency condition (\ref{}).
	
We recall that the modular invariant matrices form a group	under multiplication, which is isomorphic to the  group $G_k$ of roots of unity modulo $2k$, 
	\be 
	G_k =   \{  [\o ]_{ 2k};   \o^2 \equiv 1\  (\md 2 k) \}.\label{Gdef3}
	\ee
	Since $k$ is square-free, its prime decomposition is of the form\footnote{The function $\O (n)$ is defined as the number of distinct primes in the prime decomposition of $n$.}
$k = p_1 p_2 \ldots p_{\O (k)}$ with all $p_\r$ distinct, and 	the number of elements of $G_k$ is $ d(k) = 2^{\O (k)}$. This is also the number of independent modular invariants which we denoted by $\calm$:
\be 
\calm = |G_k|  = 2^{\O (k)}.
\ee
 The group $G_k$ is commutative and, apart from the unit element $\o_1$, all group elements have order two. A standard result \cite{Omami:2009} shows that $G_k$ is in fact isomorphic to $(\ZZ_2)^{\O (k)}$. The group $(\ZZ_2)^{\O (k)}$ is generated by $\O (k)$ `odd elements' which, under the  isomorphism correspond to 
 \be \o (p_\r), \qquad \r = 1, \ldots , \O (k).\label{oddels}
 \ee 
 For example, when $k=6$ (see (\ref{k6isom})) these odd elements are given by $\o (2) = [7]_{12} \simeq (-1,1)$ and
 $\o (3) = [5]_{12} \simeq (1,-1)$.

\subsubsection{Some representation theory}
In what follows we will need some representation theory of $G_k$. Let us first discuss its irreducible representations (irreps). Since $G_k$ is a commutative group, all the irreps are one-dimensional, and their number equals $\calm = 2^{\O (k)}$, the order of the group.  Since each group element squares to one, the irreps can  only take the values  1 or -1. 
An irrep is fully specified by giving its value   on each of the generating elements $\o (p_\r)$, leading indeed to $2^{\O (k)}$ sign choices.
 We will label the irreps as $\s^\a, \a = 1, \ldots , \calm$. For fixed $\a$, we can view $\s^\a$ as a $\calm$-dimensional  vector with components
\be 
(\s^\a)_I \equiv \s^\a (\o_I)
\ee
 taking the values 1 or -1.
%Since each group element squares to one, the entries $(\s^\a)_I $  are all either 1 or -1. 
%  We will label the irreps as $\s^\a, \a = 1, \ldots , \calm$, with 
We adopt  the convention that  $\s^1$ is the trivial representation, i.e.
\be 
(\s^1)_I = 1,\qquad  \forall I.
\ee 
The orthogonality relations between irreducible characters can be expressed as follows: 
\be 
\sum_I  (\s^\a)_I  (\s^\b)_I = \calm \d^{\a\b}, \qquad \sum_\a  (\s^\a)_I  (\s^\a)_J  = \calm \d_{IJ}.\label{charorthog}
\ee

%Since each group element squares to one, the values $\s^\a ( \o )$  the irreps take on any group element $\o$ can be either 1 or -1. 

Next we turn to the regular representation which we denote as $R$, and whose  matrices we write as $R_I \equiv R (\o_I )$.  $R$ is  the $\calm$-dimensional representation  corresponding to the  action of the group on itself
\be 
\o_I \o_J = (R_I)_J^{\ \, K} \o_K.
\ee
The representation matrices $R_I$	are symmetric, commuting, permutation matrices which square to one. %and satisfy
%\be 
% (R_I)_J^{\ K}= (R_J)_I^{\ K}.
% \ee
 A standard property for abelian groups  is that each irrep appears precisely once in the decomposition of $R$,
\be 
R = \oplus_{\a =1}^{\calm} \s^\a . \label{Rregdecomp}
\ee
The basis in which the $R$ matrices are block-diagonal according to the above decomposition  is provided by the vectors $\s^\a$. Indeed, one shows that
\be
R_I \cdot \s^\a =  \s^\a (\o_I)  \s^\a,\label{sigmaeigvect}
%\sum_K (R_I)_J^{\ K} (\s^\a)_K = \s^\a (\o_I) (\s^\a)_J
\ee
which follows from 
\be\begin{aligned}
\s^\a (\o_I) \s^\a (\o_J) &= \s^\a\left(  \sum_K  (R_I)_J^{\ K} \o_K\right) \\
&= \sum_K  (R_I)_J^{\ K} \s^\a (\o_K).
\end{aligned}\ee
A useful identity following from  (\ref{sigmaeigvect}) is that the regular representation matrices can be expressed  as
\be 
(R_I)_J^{\ \,  K} = {1\over \calm} \sum_\a \s^\a(\o_I\o_J\o_K).\label{Ritosigma}
\ee

A last representation which plays a role in our problem corresponds to  the natural action of $\o \in G_k$ on $\ZZ_{2 k}$ by multiplication, $[i]_{2 k} \to [\o i]_{2 k}$. We denote this representation by $M$ and write
$M(\o_I) \equiv M_I$. We see from (\ref{Mcomp}) that the $M_I$ are precisely the modular invariant matrices introduced before. We denote by $m^\a$ the multiplicity of the irrep $\s^\a$ in the decomposition of $M$:
\be 
 M =\oplus_{\a =1}^{\calm}  m^\a \s^\a .\label{Mdecomp}
 \ee
 Let us say a bit more about these multiplicities. It is straightforward to see that the trivial representation appears at least twice, i.e.
 \be 
 m^1 \geq 2.\label{m1ineq}
 \ee 
 Indeed, since the congruence $ [0]_{2 k}  $ is invariant under multiplication by $\o_I$, it defines an invariant subspace. A second invariant subspace is the congruence $ [k]_{2 k}  $, which is mapped to itself under multiplication by $\o_I$ since, as we established below eq. (\ref{omdef-inv-sqfree}), the $\o_I$ are odd.  In other words, we have 
 \be (M_I)^{00} =  (M_I)^{kk} =1 = (\s^1)_I, \qquad  \forall I.\label{seedtr}\ee In CFT terms this means that each modular invariant contains the (vacuum, vacuum) and the $(k,k)$ representation precisely once. As for the other multiplicities in (\ref{Dcomps}), we will see below (\ref{rhosolgen}) that they are all nonvanishing:
 \be 
 m^\a \geq 1, \qquad \forall \a.\label{mgeq0}
 \ee
 Note that the properties (\ref{m1ineq}) and (\ref{mgeq0}) imply that the representation $M$ is always bigger than the regular representation $R$.
  The decomposition (\ref{Mcomp}) implies a similar expansion for the characters, 
 $ \tr M_I  = \sum_\a m^\a (\s^\a)_I$, which can be inverted to give an formula for the multiplicities $m^\a$:
\be
 m^\a = {1 \over \calm} \sum_I    (\s^\a)_I \tr M_I.
 \ee

 \subsubsection{General formula for the ensemble weights}
 Using these properties, we can rewrite the expression (\ref{rhosgen})  for the ensemble weights  $\r^I$  in a theory with generic seed amplitude purely in terms of representation-theoretic quantities. 
  %let us first reproduce the result of \cite{} that in the exotic theories of gravity with square-free $k$ the ensemble weights $\r^I$ are all equal. 
  Recall that (\ref{rhosgen}) takes the form
 \be 
 \r_I = n \sum_J (D^{-1})_{IJ} (V^{\rm seed})_J,\label{weights2}
 \ee
 where $n$ is determined by the normalization condition that $\sum_I \r_I=1$. 
First we observe that the matrix $D_{IJ}= \tr M_I M_J$ becomes diagonal in  the  $\{ \s^\a \}$ basis:
\be \begin{aligned}
D_{IJ} &= \sum_\a m^\a (\s^\a)_I  (\s^\a)_J, \\
D_{\alpha \beta} & = \sum_{IJ} (\sigma^\alpha)_I D_{IJ}(\sigma^\beta)_J = \mathcal{M}^2 m^\alpha \delta_{\alpha\beta} ,\label{Dcomps}
\end{aligned}\ee 
where we used the orthogonality relations \eqref{charorthog}. Since $D_{IJ} $ can be shown to be invertible\footnote{ First one realizes that the matrices \({M_I}\) are linearly independent. This is because if a linear combination satisfies \(0=\sum_I a^I M_I\) then acting on \([1]_{2k}\) one gets \(0=\sum_I a^I [\omega_I]_{2k}\). Since the \(\omega_I\) are all distinct elements of \(\mathbb{Z}_{2k}\), the above is a linear combination of a sub-basis of \(\mathbb{C}^{2k}\), which leads us to conclude that \(a^I=0\), hence \(\lbrace M_I\rbrace\) is linear independent. Since \(\mathrm{tr(-^\dagger,-)}\) is a non-degenerate bilinear form on matrices, when restricted to the subspace spanned by \(\lbrace M_I\rbrace\) (which are real symmetric) it remains non-degenerate because this set is linearly independent. So \(D_{IJ}\) is non-degenerate.}, it follows that    all the multiplicities $m^\a$ are nonvanishing, as anticipated in (\ref{mgeq0}) above. 
Therefore its inverse is
\bea 
( D^{-1})_{IJ}   &=&   {1\over \calm^2} \sum_\a  (m^\a)^{-1} (\s^\a)_I  (\s^\a)_J = {1\over \calm^2} \sum_\a  (m^\a)^{-1} \s^\a (\o_I \o_J).
\eea
To proceed it will be useful to also define the components of various vectors appearing in the problem  in  
 the  $\{ \s^\a \}$ basis:  
\be 
%(V^{\rm seed})_I = {1\over \calm}\sum_\a  v^\a (\s^\a)_I. 
v^\a = \sum_I (V^{\rm seed})_I  (\s^\a)_I, \qquad \r^\a =\sum_I \r_I  (\s^\a)_I,\qquad  c^\a= \sum_I c_I  (\s^\a)_I . \label{Vdef}
\ee 
In order for the seed amplitude to lead to sensible ensemble weights satisfying $\sum_I \r_I=1$,  it   is necessary\footnote{When writing specific vector components, we use the convention that an upper index refers to a component in the $\{ \s^\a \}$-basis, e.g. $v^1 = v^{\a=1}$, while a lower index refers to the Carthesian `$I$-basis', e.g. $\r_1 = \r_{I =1}$.}  that $v^1 \neq 0$, or equivalently, $\sum_I v_I\neq 0$.
The formula (\ref{weights2}) can be simply rewritten as 
\be 
\r^\a = {v^\a \over v^1} {m^1 \over m^\a},
\ee
and converting back to the Carthesian basis we get
\be 
\r_I ={ 1 \over \calm} \sum_\a {v^\a \over v^1} {m^1 \over m^\a} (\s^\a)_I. \label{rhosolgen}
\ee 
This is our-sought-after general formula for the ensemble weights. Note that in the special case of the vacuum seed vector, we have $v^\a \propto \d^\a_1$ and we reproduce the result of \cite{Raeymaekers:2021ypf} that the ensemble weights are equal, $\r_I= \calm^{-1}, \ \forall I$.

\subsubsection{Two-boundary consistency conditions}
Now let us try to solve the consistency conditions (\ref{mastereq})  to determine the wormhole coefficients $c_I$. %for general seed vectors $v^\a$.
Due to being a highly overdetermined system, the existence if solutions will also place constraints on the allowed seed vectors.
The equations  (\ref{mastereq}) can  be written as 
\bea
\sum_{M,N} \bar c_M c_N (R_M R_N)_{IJ}&=& \sum_K D_{IK} (  \r_K \d_{KJ} - \r_K \r_J) \nonu
&=& \r_J \left( D_{IJ} -\calm {m^1 \over v^1} (V^{\rm seed})_I\right),\label{ceq3}
\eea
where in the second line we used (\ref{rhosolgen}).
Using the identity (\ref{Ritosigma}), multiplying both sides with $(\s^\a)_K$ and summing over $K$,   the system  (\ref{ceq3}) can be   reduced to
\be
\left|c^\a\right|^2 =\calm  m^\a \r_I \left( 1 -  \r^\a (\s^\a)_I  \right) , \qquad \forall \a, I .\label{ceqgen1}
\ee
Summing (\ref{ceqgen1}) over all $I$ leads to the equation
\be 
\left|c^\a \right|^2 = m^\a \left( 1- (\r^\a)^2\right), \label{csmodsq}
\ee 
which determines the coefficients $c_I$ up to phases:
\be 
c_I = {1\over \calm} \sum_\a \sqrt{m^\a} \sqrt{1 - (\r^\a)^2} e^{i \theta_\a} (\s^\a)_I.\label{cIsolgen}
\ee
We still have to check whether the remaining equations in the system (\ref{ceqgen1}) are satisfied. Substituting (\ref{csmodsq}) into  (\ref{ceqgen1}) yields
\be 
 1- (\r^\a)^2 = \calm \left(1 - (\s^\a)_I \r^\a \right) \r_I, \qquad \forall \a, I.\label{cexistencecrit}
 \ee 
This system can be viewed as a set of highly restrictive constraints on the 1-boundary seed coefficients $v^\a$. %will not have to solutions for generic $v^\a$.
 Let us now find all the solutions which make physical sense. 

Summing (\ref{cexistencecrit}) over $\a$, we obtain
\be 
 \r_I ( 1- \r_I) = { 1 \over \calm} \left(1 - \sum_J \r_J^2\right) \equiv X \qquad \forall I .\label{rhoroots}
\ee
We note that the RHS is independent  of $I$, so this equation restricts each $\r_I$ to be a root of the same quadratic equation. 

The case $\calm = 2$ is special, since (\ref{rhoroots}) is automatically satisfied for $\r_2 = (1- \r_1)$, and so is (\ref{cexistencecrit}). Therefore for $\calm =2$ the $\r_I$ are unrestricted and the general solution for the wormhole coefficients (\ref{cIsolgen}) reduces to
\be 
c_I = \sqrt{m^2 \r_1 \r_2} e^{i \theta_2} (1,-1)\label{CIMis2}
\ee

Now let us analyze the situation where $\calm >2$.
 The general solution of (\ref{rhoroots}) is determined by $\calm$ sign choices $s_I = \pm 1, I = 1, \ldots, \calm$ for the branches of  the roots of quadratic equations (\ref{rhoroots}):
 \be 
 \r_I = \half \left( 1 - s_I \sqrt{1 - 4 X}\right) .\label{rhoroots2}
 \ee
 Summing over $I$ and using (\ref{rhonorm})  we find that at least half of the $s_I$ should be  equal to 1, since
 \be 
 \sum_I s_I = { \calm- 2 \over \sqrt{1 - 4 X} } > 0 .
 \ee
  %  Imposing also that $\sum_I\r_I =1$  leads to the an expression for the  $\r_I$ in terms of  $\calm $ signs $s_I = \pm 1, \ I = 1, \ldots , \calm $ with the restriction that \comment{This assumes $\calm >2$. For $\calm =2 $ we have $s_1 = - s_2$, treat separately.}
%\be 
%\sum_I s_I \neq 0,
%\ee
Substituting into (\ref{rhoroots2}) gives
\be
\r_I = \half \left( 1 - {\calm-2 \over \sum_J s_J } s_I \right).\label{rhoconistent}
\ee
%Since $\r_I$ is invariant under an overall sign change of the $s_I$, we can restrict to the cases where
%\be 
%\sum_I s_I > 0,
%\ee
%meaning that {\em  more than half of the  $s_I$ must be equal to  plus one}.

For each choice of the $s_I$ which sum to a positive  number we should in principle check if the full set of equations (\ref{cexistencecrit}) is obeyed. This would be a daunting task in general, but in practice we can restrict attention to those cases where the  $\r_I$ in (\ref{rhoconistent}) are positive. This will reduce our work to checking  only two cases.

The first case to verify is that where all $s_I =1$. This leads to   
\begin{align}
\r_I &= {1\over \calm}, & \r^\a &= \d^\a_1, & v^\a &= v^1 \d^\a_1,
\end{align}
and it is straightforward to check that (\ref{cexistencecrit}) is obeyed. This case includes our natural starting point of  the vacuum seed amplitude, $Z_{\rm seed}^{(1)} \propto |\tilde \chi_0|^2$  but, due to  (\ref{seedtr}) one could just as well take $Z_{\rm seed}^{(1)} \propto |\tilde \chi_k|^2$.  The formula (\ref{cIsolgen}) gives the expression
for the wormhole coefficients
\be
 c_I ={1 \over \calm} \sum_{\a = 2}^\calm   \sqrt{ m^\a} e^{i \theta_\a} \s^\a (\o_I).\label{csol}
\ee

The next case to check is that where all except one of $s_I$ are equal to one, say $s_J = -1$ for some $J$. This leads to 
\begin{align}
\r_I &= \d_{IJ}, & \r^\a &= (\s^\a)_J, & v^\a &= {v^1 m^\a\over m^1} (\s^\a)_J.
\end{align}
This also trivially satisfies (\ref{cexistencecrit}). The first relation shows that there  is no ensemble and (\ref{cIsolgen}) shows that the corresponding wormhole coefficients vanish,
\be 
c_I =0, \qquad \forall I.
\ee
This class arises  if we start with a seed amplitude which is already modular invariant, $Z_{\rm seed}^{(1)}  \propto Z_J^{\rm CFT}$ for some $J$.

Next we should consider the case  where precisely two of the signs, say  $s_J, s_K$ for some $J\neq K$, are equal to minus one. Note that this requires $\calm >4$. Then the expression (\ref{rhoconistent}) gives 
\be 
\r_I = - {1\over \calm - 4} <0 , \qquad {\rm for\ } I \neq J,K.
\ee
Similarly, one shows that all remaining choices for the $s_I$ would lead to negative ensemble weights and can therefore be discarded.   

As a check on our formulas it is useful to compare them to our earlier computations in simple examples. For $k=2$,
we have
\be 
(\s^\a)_I = {\small \left( \begin{array}{cc} 1&1 \\ 1&-1\end{array}\right)}, \qquad m^\a = (3,1).
\ee
One checks that our equations (\ref{rhosolgen}) for the ensemble weight and (\ref{CIMis2}) for the wormhole coefficients in the case $\calm=2$ match with our earlier results (\ref{rhok2}) and (\ref{cIk2}).
For $k=6$, one finds
\be 
(\s^\a)_I ={\small \left( \begin{array}{cccc} 1&1& 1& 1\\ 1&-1&-1&1\\
	1&-1&1&-1\\1&-1&1&-1
\end{array}\right)}, \qquad m^\a = (6,1,2,3).
\ee
With these in hand one checks that (\ref{rhosolgen}) reproduces the ensemble weights (\ref{rhok6}), and that the formula  (\ref{csol}) for the wormhole coefficients with vacuum seed  reproduces (\ref{cIk6}).

%\subsubsection{Some lessons}
While we defer the general lessons  to be learned from this rather technical analysis to the Discussion, we will first focus on one prominent feature: the need to include somewhat exotic wormholes and their relation to topological surface defects.

\section{Exotic wormholes from surface defects}\label{Secdefects}
Both in the examples and in the general discussion, we have seen that, in sitations which describe an ensemble, the wormhole coefficients $c_I$ are typically  all nonvanishing (see e.g. (\ref{cIsolgen})).
Therefore the two-boundary seed amplitude
\be 
Z_{\mathrm{seed}}^{(2)} = |Z_{T^2\times I}|^2,\qquad 
Z_{T^2\times I}(\t_1 ,\t_2) =\sum_J c_J \chi (\t_1)^T M_J \chi (\t_2)  \label{Zseedrecap}
\ee
includes, besides the $J=1$ term  which we saw in Section (\ref{Sec:wormhole_from_canonical_qzn}) comes from the standard \emph{pure} $U(1)_k$ Chern-Simons path integral \eqref{pureCS_PI_annulus} on the wormhole geometry, also other terms which represent some kind of  exotic wormhole contributions.
These 
% In the examples which do describe an ensemble, the  pure Chern-Simons amplitude (\ref{}) is only part of the answer, and one needs to include additional `exotic wormhole' contributions corresponding to
 correspond to  2-boundary amplitudes seed constructed from nondiagonal modular invariants. As we shall presently describe, these have a nice bulk interpretation in terms of path integrals including  topological surface defects.
%
%In the previous section we argued that, generically,  the bulk dual
%theory dual to an ensemble of rational boson CFTs is required to have a  seed amplitude on the torus times an interval  of the form
%
%\be 
%Z_{\mathrm{seed}}^{(2)} = |Z_{T^2\times I}|^2,\qquad 
%Z_{T^2\times I}(\t_1 ,\t_2) = c^J \chi (\t_1)^T M_J \chi (\t_2)  .\label{Zseedrecap}
%\ee
%While the term with $J=1$ corresponds to the standard \emph{pure} $U(1)_k$ Chern-Simons path integral \eqref{pureCS_PI_annulus} on the wormhole geometry, the remaining terms represent some kind of  `exotic' wormhole contributions. In this section, we identify these contributions   as coming  from a Chern-Simons path integral on the torus times an interval which include  a nontrivial topological surface defect. 
These defects were studied by Kapustin and Saulina in \cite{Kapustin:2010if} (see also  \cite{Fuchs:2002cm,Kapustin:2010hk}), and were recently discussed from a somewhat different perspective in \cite{Roumpedakis_2023}. 
Far from doing justice to the subject of surface defects in topological field theories and the  mathematical structure of  monoidal 2-categories underlying them,  we will here only review their construction in the context which is relevant for our discussion and perform some consistency checks.

Let us recall from Section \ref{Sec:wormhole_from_canonical_qzn} that quantization of the pure \(U(1)\) theory on the annulus leads to the Hilbert space \eqref{Hilb-sp_annulus} and partition function \eqref{pureCS_PI_annulus}, which in turn can be interpreted as a path integral on the wormhole geometry \(T^2\times I\) upon identifying the extra dimension with a compactified Euclidean time. There, the quantum number labelling representations and characters of the chiral algebra \(\mathcal{A}_k\) is \(i\in \mathbb{Z}_{2k}\).\footnote{\(i\) never denotes the imaginary unit in this Section.} On the wormhole geometry there are two chiral algebras associated to the two boundaries and from the partition function \eqref{pureCS_PI_annulus} we see that their representations pair diagonally. Instead, from the component expression \eqref{Mcomp} we see that in each term with \(J\neq 1\) in \eqref{Zseedrecap} the representation of charge \(i\) on one boundary pairs with the representation of charge \(\omega_J i\) (modulo \(2k/\alpha_{J}\), and if both \(i\) and \(j\) are divisible by \(\alpha_J\)) on the other boundary.

To justify the introduction of surface defects, let us also recall that the quantum number \(i\) is related geometrically to the holonomy of the bulk gauge field along the non contractible cycle on the annulus. More specifically, \(\frac{k}{\pi}\oint A = i\mod 2k\) as explained in \cite{Elitzur:1989nr}.\footnote{To compare with their expressions, notice that the normalization of the level differs from our convention by a factor of 2.} {   {This holonomy} is a dynamical coordinate on the phase space of the theory  {which has to be summed over in the path integral.}  Consequently, from a path integral point of view it is somewhat intuitive to geometrically engineer the modular invariant \(M_J\) as follows (see Figure\ \ref{Fig:surface_defect}): one inserts a surface defect \(\mathcal{S}_J\) inside the wormhole geometry, parallel to the boundaries, where the holonomy of the bulk gauge field has a discontinuity subject to the sewing conditions 
\begin{figure}
\centering
    \begin{subfigure}[t]{.5\textwidth}
       \centering
       \includegraphics[width=1\textwidth]{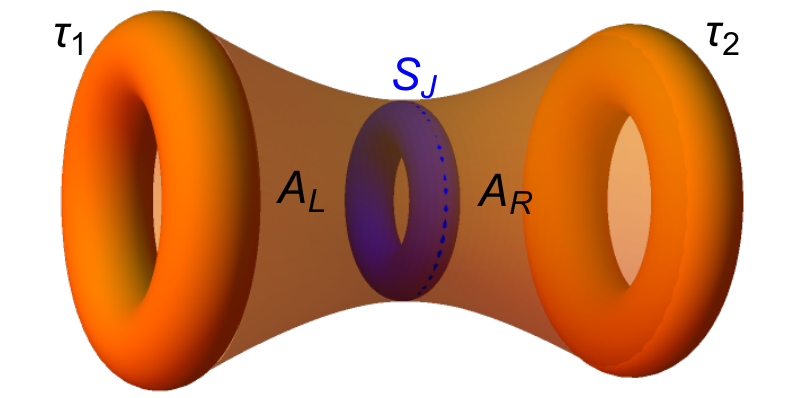}     
       \caption{}
    \end{subfigure}\quad
    \begin{subfigure}[t]{.26\textwidth}
       \centering
       \includegraphics[width=1\textwidth]{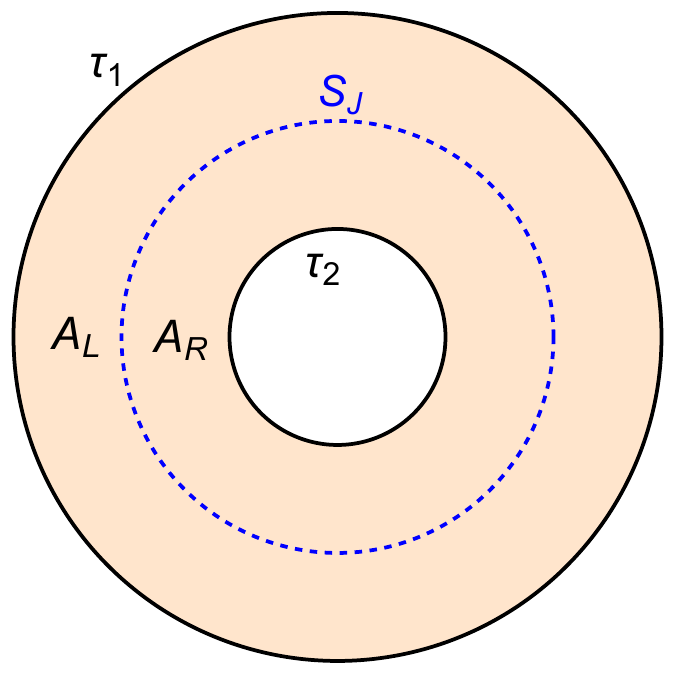}
       \caption{}
    \end{subfigure}
\caption{{Wormhole geometry \(T^2\times I\) with inclusion of the surface defect \(\mathcal{S}_J\) along the non trivial 2-cycle in the bulk. In (a) the \(I\) radial factor is displayed as connecting the two boundary components. In (b) only the \lq spatial\rq\ annulus slice is displayed (the compact \lq Euclidean time\rq\  direction is perpendicular to the paper). \(\t_1\) and \(\t_2\) denote the two boundary complex moduli, while \(A_L\) and \(A_R\) are the gauge fields on the two sides of the defect.}}
\label{Fig:surface_defect}
\end{figure} 
\be
\left(\frac{k}{\pi}\oint A_{L}\right)-\omega_{J} \left(\frac{k}{\pi} \oint A_{R}\right) = 0\mod\frac{2k}{\alpha_J} 
\ee
{whenever \(i_{L,R}:=\frac{k}{\pi}\oint A_{L,R}\) are divisible by \(\alpha_J\) (here \(A_{L,R}\) are the gauge fields on the two sides of \(\mathcal{S}_J\))}. In terms of the divisor \(\delta_{J}\) associated to \(\omega_{J}\), this can be equivalently rewritten as\footnote{The relation can be seen as follows. Let us call \(i_{L,R} = \frac{k}{\pi} \oint A_{L,R}\) and assume \(i_L = \omega i_R \mod(2k/\alpha)\). Now using definition \eqref{omdef} and adding and subtracting \(\omega\) we get 
\[ i_L - i_R = (\omega -1) i_R + \frac{2k}{\alpha} N = \frac{2k}{\delta}  \frac{i_R}{\alpha} s + \frac{2k}{\alpha}\tilde{N} \]
for some integers \(N,\tilde{N}\). By the relation of \(\delta\) and \(\alpha = \gcd(\delta, k/\delta)\) we see that \(i_L - i_R \in (2k/\delta)\mathbb{Z}\). Analogously one proves that \(i_L + i_R \in (2\delta)\mathbb{Z}\). On the other hand, starting from the two latter relations, one knows that there exist integers \(a,b\) such that \(i_L = \delta a + \frac{k}{\delta} b\) and \(i_R = \delta a - \frac{k}{\delta} b\). { It is immediate to see that both \(i_{L,R}\) are divisible by \(\alpha\), while it is more tedious} to check that this implies \(\omega i_R = i_L \mod (2k/\alpha)\).}
\be
\frac{k}{\pi}\oint(A_{L}-A_{R})|_{\mathcal{S}}=0\mod\frac{2k}{\delta_{J}},\qquad\frac{k}{\pi}\oint(A_{L}+A_{R})|_{\mathcal{S}}=0\mod2\delta_{J}. \label{sewing_cond_surface_def}
\ee
Notice that since the sewing conditions are topological they do not break diffeomorphism invariance, hence their insertions produce quantities which are invariant under diagonal modular transformations, by the same argument as given below \eqref{whmodprop}. Moreover when \(\omega_{J}\) is taken to be the identity 
%(corresponding to picking the diagonal modular invariant in \eqref{bilinear}),
we are describing a trivial or invisible defect, and the path integral agrees with the bilinear diagonal modular invariant \eqref{pureCS_PI_annulus}. 

The surface \(\mathcal{S}_J\) with sewing conditions enforcing \eqref{sewing_cond_surface_def} can be precisely identified as a topological defect of the type introduced in \cite{Kapustin:2010if}. Their full analysis from a Lagrangian point of view requires some more care than what we explained above, we refer also to \cite{Roumpedakis_2023} for a recent detailed review.\footnote{In particular the sewing conditions \eqref{sewing_cond_surface_def} follow directly from Eq.\ (6.17) in \cite{Roumpedakis_2023}, which is taken as a definition of the defect.} It is now clear that the terms with \(J\neq 1\) in the seed amplitude (\ref{Zseedrecap}) arise from path integrals with insertions of nontrivial surface defects.

For completeness we notice that these surfaces could alternatively be introduced using the so called \lq folding trick\rq\ (see Figure  \ref{Fig:surface_defect_folded}): 
\begin{figure}
\centering
\includegraphics[width=0.4\textwidth]{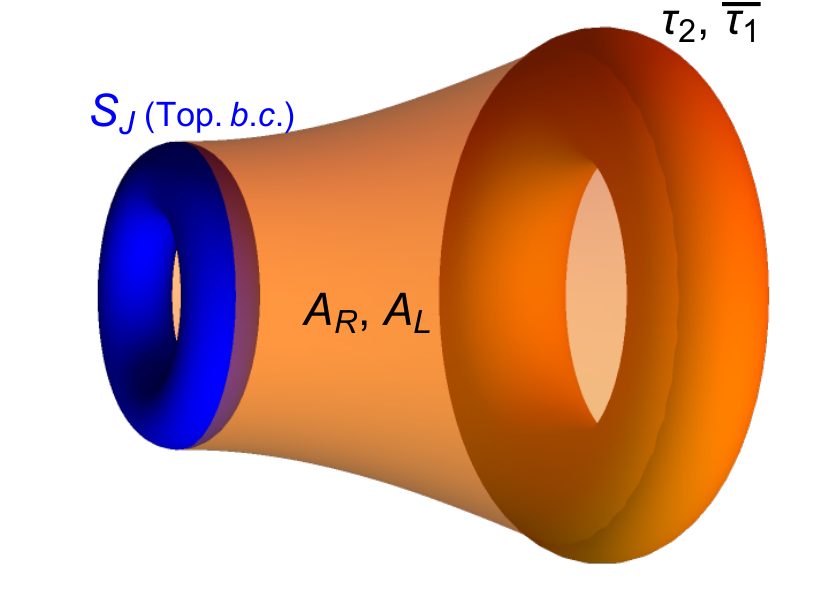}
\caption{{  Folding trick. The wormhole geometry of Fig.\ \ref{Fig:surface_defect}.a has been \lq folded\rq\ along the surface defect, which now behaves as a topological boundary condition. Due to orientation reversing one of the two sides, the theory has gauge group \(U(1)_k\times U(1)_{-k}\) and one of the boundary conditions has been conjugated.}}
\label{Fig:surface_defect_folded}
\end{figure}
one considers a \lq folded\rq\ \(U(1)_{k}\times U(1)_{-k}\) Chern-Simons theory on the torus times an interval \([0,1]\) whith dynamical boundary conditions on \(T^{2}\times\{0\}\) fixing modular parameters \(\tau_{1}\) and \(\tau_{2}\) for the two gauge fields, and topological boundary conditions on \(T^{2}\times\{1\}\) that essentially relate the holonomies of the two gauge fields as in \eqref{sewing_cond_surface_def} (up to orientation). The path integral then picks a state on the Hilbert space at \(T^{2}\times\{0\}\), which is invariant under large diffeomorphisms (diagonal modular transformations on \(\tau_{1},\tau_{2}\)) because the boundary conditions at \(T^{2}\times\{1\}\) are topological. Reversing the orientation of one of the two \(U(1)\) theories makes the above construction equivalent to the \lq unfolded\rq\ \(U(1)_{k}\) theory on \(T^{2}\times[0,2]\), where the two opposite boundaries have dynamical boundary conditions and the topological boundary conditions at \(S=T^{2}\times\{1\}\) are interpreted as sewing conditions on the topological defect.

On general grounds, two topological surface defects can be fused together to form a new defect. This fusion gives the set of defects the structure of a monoid, since the inverse is not guaranteed to exist, with the trivial defect playing the role of a unit element. The particular fusion rule found in \cite{Kapustin:2010if} is
\be\begin{aligned}
\mathcal{S}_{I} \circ \mathcal{S}_{J} & = \gcd \left(\d_I, \d_J, \frac{k}{\d_I}, \frac{k}{\d_J}\right)\mathcal{S}_{K(I,J)} , \\ 
\text{where}\quad \d_{K(I,J)} & := \mathrm{lcm}\left(\gcd\left(\d_I, \frac{k}{\d_J}\right), \gcd \left(\d_J,\frac{k}{\d_I}\right)\right).
\end{aligned}\ee
From the above considerations, the fusion of the defects should agree with the monoidal structure \eqref{Mmultgen} of the modular invariant matrices \(M_J\) under multiplication. It is a heartening consistency check to indeed verify this, as we do in Appendix \ref{app:fusion-rules-check}.

\section{Discussion}
In this paper we have analyzed the implications of the requirement that the 2-boundary  bulk   amplitude should capture the variance of the ensemble distribution, under the plausible assumptions  that it comes from a seed amplitude $Z^{(1)}_{\rm seed}$ which is holomorphically factorized (\ref{holfact}) and satisfies the Cotler-Jensen property (\ref{whmodprop}). We focused on ensembles of rational compact boson CFTs, and were able to analyze the problem in general when the level $k$ is square-free. 

\subsection{Some lessons}
Let us summarize the results of this analysis and some lessons we can draw from them. When the ensemble consists of more than two CFTs, i.e. when the level $k$ is not a prime number, we found that the % constraiExcept in the case of prime $k$, where there are only two modular invariants, we saw that the
 consistency constraints (\ref{mastereq}) are highly constraining and allow for only  two possibilities.

 A first and somewhat trivial possibility  occurs when we start with a 1-boundary seed amplitude which is already  modular invariant by itself; this picks out a specific CFT and, as here is no ensemble, the wormhole coefficients $c_I$  all vanish.  This can be thought of as starting from a UV-complete complete theory and illustrates how the ensemble averaging phenomenon is related to effective gravity theories which are not UV complete. 
 This situation is similar to e.g. tensionless string theory on AdS$_3$, which has a seed amplitude which is already modular invariant \cite{Eberhardt:2021jvj}. As discussed in \cite{Benini:2022hzx}, starting from a  modular invariant seed amplitude can also interpreted as   gauging a non-anomalous global 1-form symmetry, and is consistent with the expectation that global symmetries  should be absent in a UV-complete theory of gravity.
 
 The second possibility leads to an  ensemble with maximal  entropy and uniform ensemble weights. %, and the only one which leads to a genuine ensemble, 
 %arises from the most  natural starting point where the 1-boundary seed amplitude is simply the pure Chern-Simons path-integral on the solid torus.
 % In fact, for square-free $k$, the only allowed possibility leads to the maximal  entropy ensemble with uniform ensemble weights. 
  It is reassuring that this possibility also arises from the most  natural starting point, where the 1-boundary seed amplitude is simply the pure Chern-Simons path-integral on the solid torus. Interestingly, the same ensemble could be obtained from the charge-$k$ excited state\footnote{One  reason  that the vacuum and the charge-$k$ excited state  share certain properties is that they are `related by half a unit of spectral flow'. Spectral flow by an amount $a$ is an automorphism acting as $L_n'= L_n +2 \sqrt{k} a J_n + 2 k a^2\d_{n,0},\ J_n' = J_n + 2\sqrt{k} a \d_{n,0}$ under which charges are shifted as $i \to i + 2 k a$. For $a= 1/2$ the charge lattice is preserved and the vacuum and charge $k$ state are interchanged. } $\tilde \chi_k$.

  Modulo this last subtlety, it is interesting that the  consistent starting points are in some sense either
%It is interesting that only these two starting points, either a 
fully UV complete   or a maximally UV-ignorant, containing only the vacuum, % exhaust  all  possibilities,
 and that intermediate starting points, such as a $Z^{(1)}_{\rm seed}$ containing the vacuum plus a few additional matter representations, are not consistent.

Another conclusion we  reached is that in none of the examined cases the wormhole coefficients agree with the naive answer (\ref{cpureCS}) given by the pure Chern-Simons path integral. In the models which  describe an ensemble, we found that it is always necessary to include additional contributions from exotic wormholes, which are realized as a  path integral in the presence of a Kapustin-Saulina surface defect.
In the examples  describing a single CFT, which include the $k=1$ case and the UV-complete theories, the wormhole coefficients vanish. This means that in these cases, we are instructed not to include two-boundary geometries in the path integral, lest we get an inconsistent result. This is perhaps similar to  string theory on AdS$_5 \times$S$^5$, where wormhole saddle points exist in the low-energy approximation \cite{Maldacena:2004rf}  which should most likely  not be included in the path integral.

\subsection{Outlook}
Let us end by discussing some open problems and possible generalizations. Perhaps the biggest question raised by our analysis is if the 2-boundary seed amplitude, derived here  from requiring a consistent holographic  description of the ensemble, can be given an intrinsic three-dimensional geometric meaning. Our arguments showed that the two-boundary path integral should be performed with the insertion of a formal combination  $\oplus_I ( c_I \cals_I)$ of surface defects, and it would be interesting if this prescription could be given a physical interpretation, for example as arising from gauging a certain global symmetry.   Such insights would go a long way to generalize our work to determining the   seed amplitudes for multi-boundary amplitudes and higher genus boundaries. It will be interesting to see if the uniform ensemble distribution arising from the vacuum seed will pass the consistency tests arising  on these topologies as well. 

An obvious question is how our work generalizes to the holographic description of ensembles in other rational CFTs. A first class to consider would be the rational boson ensembles in which $k$ has nontrivial square divisors. As we saw above, the additional complications stem, on the boundary side, from the fact that the chiral algebra extends or,  in the bulk, from the existence  of non-invertible zero-form symmetries associated to certain surface defects.
We briefly touched on this class of theories in the suggestive example in par. \ref{exk9}, and a general analysis is in progress \cite{Paolo}. Also of interest would be the generalization to multiple compact bosons in rational Narain theories  (see \cite{Furuta:2023xwl} for a classification) and to Wess-Zumino-Witten models as well as  Virasoro minimal models. As a by-product it would be interesting to  harness the existing  knowledge of modular invariants through their ADE classification \cite{Cappelli:1987xt} to holographically determine or verify the fusion rules of surface defects in the associated Chern-Simons theories. In the interest of understanding the Euclidean path integral in more general contexts like \cite{Penington:2019kki,Almheiri:2019qdq},  it would also be desirable give a  clear bulk path-integral derivation of the partition function on the torus times an interval with the insertion of a surface defect,
  following related recent studies \cite{Porrati:2019knx,Porrati:2021sdc}.  
  
In order to progress towards  the ultimate goal of understanding the ensemble interpretation of pure 3D gravity, we should extend the analysis of multi-boundary consistency constraints to ensembles of irrational CFTs.  A natural first step  would be to consider the Narain ensembles of irrational free boson theories \cite{Afkhami-Jeddi:2020ezh,Maloney:2020nni}.    In this case, the required Poincar\'e-summed `exotic wormhole' contributions were determined in \cite{Collier:2021rsn}, and it would be of interest to see if those  arise form a sensible seed amplitude and  allow for an interpretation in terms of surface defects. The current work may also shed light on the ensemble description of pure gravity at certain negative values of the central charge \cite{Raeymaekers:2020gtz,Campoleoni:2017xyl}, where the chiral algebra has been argued to be  a twisted version of the rational boson  algebra $\cala_k$.

 \section*{Acknowledgements}
This work was supported
by the Grant Agency of the Czech Republic under the grant EXPRO 20-25775X.
%\section{Wormhole partition functions: solution of consistency conditions}

%\section{Missing wormholes and surface defects}

\begin{appendix}

\section{Fusion rules of surface defects }  
\label{app:fusion-rules-check}
The surface defects of \cite{Kapustin:2010if} are parametrized by the divisors of the level \(k\). Given two divisors \(\d_I , \d_J\), the fusion of the associated defects \(\mathcal{S}_I\) and \(\mathcal{S}_J\) is governed by the rule\footnote{In this section we use the notation \(G(-,-)\) and \(L(-,-)\) for the greatest common divisor and lowest common multiple, respectively.} 
\be
\mathcal{S}_{I} \circ \mathcal{S}_{J} = G\left(\d_I, \d_J, \frac{k}{\d_I}, \frac{k}{\d_J}\right)\mathcal{S}_{K(I,J)} ,\qquad \d_{K(I,J)}:=L\left(G\left(\d_I, \frac{k}{\d_J}\right),G\left(\d_J,\frac{k}{\d_I}\right)\right). \label{eq: fusion-KS-app}
\ee
In this Section we show the agreement between this formula and the multiplication rule \eqref{Mmultgen} satisfied by the modular invariant matrices \(M_I\).

First of all we notice that, by definition of \(\alpha_I =G(\d_I , k/ \d_I )\) and associativity of gcd, the prefactors agree:
\be 
G\left(\d_I, \d_J, \frac{k}{\d_I}, \frac{k}{\d_J}\right) = G\left(\a_I ,\a_J\right).
\ee

Now we will rewrite the formula for \(\d_{K(I,J)}\) in a more suitable form. Using standard properties of gcd and lcm one can show that
\be
G(\d_I,\d_J) G\left(\d_I,\frac{k}{\d_J}\right) = \d_I G\left(\d_I, \d_J ,\frac{k}{L(\d_I ,\d_J )}\right) = \d_I G(\alpha_I,\a_J).
\ee
Thus, collecting \(\frac{G(\alpha_I,\a_J)}{G(\d_I ,\d_J)}\), \eqref{eq: fusion-KS-app} can be rewritten as\footnote{We notice that this is consistent with the analogous result of \cite{Roumpedakis_2023} at genus \(g=1\).}
\be 
\d_{K(I,J)}=\frac{G(\a_I , \a_J) L(\d_I ,\d_J )}{G(\d_I ,\d_J)}.
\ee

To compare with the multiplication law (\ref{ommult}), we must show that
\be 
\d_{K(I,J)} \stackrel{?}{=} \delta(\omega_I \bullet \omega_J) = L(\a_I , \a_J) G\left( \frac{\omega_I \omega_J + 1}{2}, \frac{k}{L(\a_I , \a_J)^2} \right),
\ee
where we used \eqref{omdef-inv} on the RHS.

Starting from the RHS, we multiply and divide by $G(\d_I ,\d_J )$, use distributivity and associativity of the gcd and replace $\d_I = \d (\o_I) =\a_I  G\left(\frac{\o_I +1}{2},\frac{k}{\a_I ^{2}}\right)$
to get
\begin{equation}
\begin{aligned} & \d(\o_I \bullet \o_J)  =\frac{L(\a_I ,\a_J )}{G(\d_I ,\d_J )}G\left(\d_I \frac{\o_I \o_J +1}{2},\d_J \frac{\o_I \o_J +1}{2},\d_I \frac{k}{L(\a_I ,\a_J )^{2}},\d_J \frac{k}{L(\a_I ,\a_J )^{2}}\right)\\
 & =\frac{L(\a_I ,\a_J )}{G(\d_I ,\d_J )} G \Bigg(\a_I \frac{\o_I +1}{2}\frac{\o_I \o_J +1}{2},\a_J \frac{\o_J +1}{2}\frac{\o_I \o_J +1}{2}, \\
 & \qquad \underbrace{\frac{k}{L(\a_I ,\a_J )}\frac{\o_I \o_J +1}{2},\a_I \frac{\o_I +1}{2}\frac{k}{L(\a_I ,\a_J )^{2}},\a_J \frac{\o_J +1}{2}\frac{k}{L(\a_I ,\a_J )^{2}},\frac{k^{2}}{L(\a_I ,\a_J )^{3}}}_{(*)} \Bigg).
\end{aligned}
\end{equation}
Let us focus on \((*)\). Rewriting \(\frac{\o_I \o_J +1}{2}= \o_I\frac{\o_J +1}{2}+\frac{1-\o_J}{2}\), noticing that \(\a_I \frac{\o_I +1}{2}\) and repeatedly using standard properties of the gcd one can see that 
\be
(*) = G\left(\frac{k}{L(\a_I ,\a_J )},\a_I \frac{\o_I +1}{2}\frac{k}{L(\a_I ,\a_J )^{2}},\a_J \frac{\o_J +1}{2}\frac{k}{L(\a_I ,\a_J )^{2}}\right).
\ee

%Here we rewrite \(\frac{\o_I \o_J +1}{2}= \o_I\frac{\o_J +1}{2}+\frac{1-\o_J}{2}\), use associativity and the modular property $G(A+mB,B)=G(A,B)$ to simplify it to
%\begin{equation}
%\begin{aligned}(*) & =\frac{k}{L(\a_I \a_J )^{2}}G\left(\cancel{L(\a_I ,\a_J )\o_I \frac{\o_J +1}{2}}+L(\a_I \a_J )\frac{1-\o_I }{2},\a_I \frac{\o_I +1}{2},\a_J \frac{\o_J +1}{2},\frac{k}{L(\a_I ,\a_J )}\right) \\
% & =\frac{k}{L(\a_I \a_J )^{2}}G\left(L(\a_I ,\a_J )\cancel{\frac{1-\o_I }{2}},\a_I \frac{\o_I +1}{2},\a_J \frac{\o_J +1}{2},\frac{k}{L(\a_I ,\a_J )}\right) \\
% & =\frac{k}{L(\a_I \a_J )^{2}}G\left(\underbrace{L(\a_I ,\a_J )},\a_I \frac{\o_I +1}{2},\a_J \frac{\o_J +1}{2},\underbrace{\frac{k}{L(\a_I ,\a_J )}}\right) \\
% & =G\left(\frac{k}{L(\a_I ,\a_J )},\a_I \frac{\o_I +1}{2}\frac{k}{L(\a_I ,\a_J )^{2}},\a_J \frac{\o_J +1}{2}\frac{k}{L(\a_I ,\a_J )^{2}}\right).
%\end{aligned}
%\end{equation}
%In the first line we used the modular property with respect to \(\a_I \frac{\o_I +1}{2}\); in the second line we used that \(G\left(\frac{\o+1}{2},\frac{\o-1}{2}\right)=1\); in the third line we noticed that \(L(\a_I, \a_J)\) divides \(\frac{k}{L(\a_I,\a_J)}\).

Notice that $\a_I \frac{\o_I +1}{2}\frac{\o_I \o_J +1}{2}=\a_I \frac{\o_I +1}{2}\frac{\o_J +1}{2}+2\o_J M\frac{k}{\a_I }$
for some integer $M$ satisfying $\o_I ^{2}=1+M(4k/\a_I ^{2})$
(and analogously for $\o_J $). Since in the main expression
we work modulo $\frac{k}{L(\a_I ,\a_J )}$, we can substitute
$\a_I \frac{\o_I +1}{2}\frac{\o_I \o_J +1}{2}\sim\a_I \frac{\o_I +1}{2}\frac{\o_J +1}{2}$:
\begin{equation}
\begin{aligned}\d_I (\o_I \bullet\o_J ) & =\frac{L(\a_I ,\a_J )}{G(\d_I ,\d_J )} G \Bigg(G(\a_I ,\a_J )\frac{\o_I +1}{2}\frac{\o_J +1}{2},\frac{k}{L(\a_I ,\a_J )}, \\ 
 & \qquad\qquad\qquad\qquad\qquad \a_I \frac{\o_I +1}{2}\frac{k}{L(\a_I ,\a_J )^{2}},\a_J \frac{\o_J +1}{2}\frac{k}{L(\a_I ,\a_J )^{2}}\Bigg)\\
 & =\frac{1}{G(\d_I ,\d_J )}G\left(\a_I \frac{\o_I +1}{2}\a_J \frac{\o_J +1}{2},k,\a_I \frac{\o_I +1}{2}\frac{k}{L(\a_I ,\a_J )},\a_J \frac{\o_J +1}{2}\frac{k}{L(\a_I ,\a_J )}\right)
\end{aligned}
\end{equation}
where we used $G(\a_I ,\a_J )L(\a_I ,\a_J )=\a_I \a_J $.
Now using the definition of $\d (\o )$ we substitute $\a \frac{\o +1}{2}=R \d $, where $R $ is an integer satisfying $G(R ,k/\a )=G(R ,\a )=G(R ,k)=1$. Since under the gcd we can work modulo \(k\), both \(R_I\) and \(R_J\) can be simplified, leading to
\begin{equation}
\begin{aligned} & =\frac{1}{G(\d_I ,\d_J )} G \Bigg(\underbrace{\d_I \d_J ,k}_{=L(\d_I,\d_J)G(\a_I,\a_J)} ,\underbrace{\frac{k}{L(\a_I ,\a_J )}\d_I ,\frac{k}{L(\a_I ,\a_J )}\d_J }_{=\frac{k}{L(\a_I ,\a_J )}G(\d_I ,\d_J )}\Bigg) \\
 & =\frac{G(\a_I ,\a_J )}{G(\d_I ,\d_J )}G\left(L(\d_I ,\d_J ),\frac{k}{\a_I \a_J }G(\d_I ,\d_J )\right).
\end{aligned}
\end{equation}
Using prime decomposition we will check below that $L(\d_I ,\d_J )$ divides $\frac{k}{\a_I \a_J }G(\d_I ,\d_J )$, hence the final result is
\begin{equation}
\d_I (\o_I \bullet\o_J )=\frac{G(\a_I ,\a_J )L(\d_I ,\d_J )}{G(\d_I ,\d_J )} = \d_{K(I,J)}
\end{equation}
as desired.

Let us check the promised relation $G\left(L(\d_I ,\d_J ),\frac{k}{\a_I \a_J }G(\d_I ,\d_J )\right)=L(\d_I ,\d_J )$.
Fix the prime decompositons
\begin{equation}
k=\prod_{j}p_{j}^{n_{j}},\qquad \d_I =\prod_{j}p_{j}^{m_{j}},\qquad\d_J =\prod_{j}p_{j}^{\tilde{m}_{j}},
\end{equation}
with $m_{j},\tilde{m}_{j}\leq n_{j}$. Then $\a_I =\prod_{j}p_{j}^{\min(m_{j},n_{j}-m_{j})}$,
$\a_J =\prod_{j}p_{j}^{\min(\tilde{m}_{j},n_{j}-\tilde{m}_{j})}$
and $L(\d_I ,\d_J )=\prod_{j}p_{j}^{\max(m_{j},\tilde{m}_{j})}$.
Also the other combination in the gcd is
\begin{equation}
\frac{k}{\a_I \a_J }G(\d_I ,\d_J )=\prod_{j}p_{j}^{n_{j}+\min(m_{j},\tilde{m}_{j})-\min(m_{j},n_{j}-m_{j})-\min(\tilde{m}_{j},n_{j}-\tilde{m}_{j})}.
\end{equation}
For each $j$, the exponents in this last expression can be explicitly seen to satisfy
\begin{equation}
n+\min(m,\tilde{m})-\min(m,n-m)-\min(\tilde{m},n-\tilde{m}) \geq\begin{cases}
m & m>\tilde{m}\\
\tilde{m} & m<\tilde{m}
\end{cases}.
\end{equation}
The RHS is the form of the exponents of $L(\d_I ,\d_J )$, hence the
latter divides $\frac{k}{\a_I \a_J }G(\d_I ,\d_J )$ as
claimed.

\end{appendix}                                                                                                                                                                       

\bibliographystyle{ytphys}
\bibliography{refsolitons}

\providecommand{\href}[2]{#2}\begingroup\raggedright\begin{thebibliography}{10}

\bibitem{Penington:2019kki}
G.~Penington, S.~H. Shenker, D.~Stanford, and Z.~Yang, ``{Replica wormholes and
  the black hole interior},''
  \href{http://dx.doi.org/10.1007/JHEP03(2022)205}{{\em JHEP} {\bfseries 03}
  (2022) 205}, \href{http://arxiv.org/abs/1911.11977}{{\ttfamily
  arXiv:1911.11977 [hep-th]}}.

\bibitem{Almheiri:2019qdq}
A.~Almheiri, T.~Hartman, J.~Maldacena, E.~Shaghoulian, and A.~Tajdini,
  ``{Replica Wormholes and the Entropy of Hawking Radiation},''
  \href{http://dx.doi.org/10.1007/JHEP05(2020)013}{{\em JHEP} {\bfseries 05}
  (2020) 013}, \href{http://arxiv.org/abs/1911.12333}{{\ttfamily
  arXiv:1911.12333 [hep-th]}}.

\bibitem{Saad:2019lba}
P.~Saad, S.~H. Shenker, and D.~Stanford, ``{JT gravity as a matrix integral},''
  \href{http://arxiv.org/abs/1903.11115}{{\ttfamily arXiv:1903.11115
  [hep-th]}}.

\bibitem{Maloney:2007ud}
A.~Maloney and E.~Witten, ``{Quantum Gravity Partition Functions in Three
  Dimensions},'' \href{http://dx.doi.org/10.1007/JHEP02(2010)029}{{\em JHEP}
  {\bfseries 02} (2010) 029},
\href{http://arxiv.org/abs/0712.0155}{{\ttfamily arXiv:0712.0155 [hep-th]}}.
%%CITATION = ARXIV:0712.0155;%%.

\bibitem{Cotler:2020ugk}
J.~Cotler and K.~Jensen, ``{AdS$_{3}$ gravity and random CFT},''
  \href{http://dx.doi.org/10.1007/JHEP04(2021)033}{{\em JHEP} {\bfseries 04}
  (2021) 033}, \href{http://arxiv.org/abs/2006.08648}{{\ttfamily
  arXiv:2006.08648 [hep-th]}}.

\bibitem{Keller:2014xba}
C.~A. Keller and A.~Maloney, ``{Poincare Series, 3D Gravity and CFT
  Spectroscopy},'' \href{http://dx.doi.org/10.1007/JHEP02(2015)080}{{\em JHEP}
  {\bfseries 02} (2015) 080}, \href{http://arxiv.org/abs/1407.6008}{{\ttfamily
  arXiv:1407.6008 [hep-th]}}.

\bibitem{Benjamin:2019stq}
N.~Benjamin, H.~Ooguri, S.-H. Shao, and Y.~Wang, ``{Light-cone modular
  bootstrap and pure gravity},''
  \href{http://dx.doi.org/10.1103/PhysRevD.100.066029}{{\em Phys. Rev. D}
  {\bfseries 100} no.~6, (2019) 066029},
  \href{http://arxiv.org/abs/1906.04184}{{\ttfamily arXiv:1906.04184
  [hep-th]}}.

\bibitem{Benjamin:2020mfz}
N.~Benjamin, S.~Collier, and A.~Maloney, ``{Pure Gravity and Conical
  Defects},'' \href{http://dx.doi.org/10.1007/JHEP09(2020)034}{{\em JHEP}
  {\bfseries 09} (2020) 034}, \href{http://arxiv.org/abs/2004.14428}{{\ttfamily
  arXiv:2004.14428 [hep-th]}}.

\bibitem{Maxfield:2020ale}
H.~Maxfield and G.~J. Turiaci, ``{The path integral of 3D gravity near
  extremality; or, JT gravity with defects as a matrix integral},''
  \href{http://dx.doi.org/10.1007/JHEP01(2021)118}{{\em JHEP} {\bfseries 01}
  (2021) 118}, \href{http://arxiv.org/abs/2006.11317}{{\ttfamily
  arXiv:2006.11317 [hep-th]}}.

\bibitem{Afkhami-Jeddi:2020ezh}
N.~Afkhami-Jeddi, H.~Cohn, T.~Hartman, and A.~Tajdini, ``{Free partition
  functions and an averaged holographic duality},''
  \href{http://dx.doi.org/10.1007/JHEP01(2021)130}{{\em JHEP} {\bfseries 01}
  (2021) 130}, \href{http://arxiv.org/abs/2006.04839}{{\ttfamily
  arXiv:2006.04839 [hep-th]}}.

\bibitem{Maloney:2020nni}
A.~Maloney and E.~Witten, ``{Averaging over Narain moduli space},''
  \href{http://dx.doi.org/10.1007/JHEP10(2020)187}{{\em JHEP} {\bfseries 10}
  (2020) 187}, \href{http://arxiv.org/abs/2006.04855}{{\ttfamily
  arXiv:2006.04855 [hep-th]}}.

\bibitem{Perez:2020klz}
A.~P\'erez and R.~Troncoso, ``{Gravitational dual of averaged free
  CFT\textquoteright{}s over the Narain lattice},''
  \href{http://dx.doi.org/10.1007/JHEP11(2020)015}{{\em JHEP} {\bfseries 11}
  (2020) 015}, \href{http://arxiv.org/abs/2006.08216}{{\ttfamily
  arXiv:2006.08216 [hep-th]}}.

\bibitem{Dymarsky:2020pzc}
A.~Dymarsky and A.~Shapere, ``{Comments on the holographic description of
  Narain theories},'' \href{http://dx.doi.org/10.1007/JHEP10(2021)197}{{\em
  JHEP} {\bfseries 10} (2021) 197},
  \href{http://arxiv.org/abs/2012.15830}{{\ttfamily arXiv:2012.15830
  [hep-th]}}.

\bibitem{Datta:2021ftn}
S.~Datta, S.~Duary, P.~Kraus, P.~Maity, and A.~Maloney, ``{Adding flavor to the
  Narain ensemble},'' \href{http://dx.doi.org/10.1007/JHEP05(2022)090}{{\em
  JHEP} {\bfseries 05} (2022) 090},
  \href{http://arxiv.org/abs/2102.12509}{{\ttfamily arXiv:2102.12509
  [hep-th]}}.

\bibitem{Benjamin:2021wzr}
N.~Benjamin, C.~A. Keller, H.~Ooguri, and I.~G. Zadeh, ``{Narain to Narnia},''
  \href{http://dx.doi.org/10.1007/s00220-021-04211-x}{{\em Commun. Math. Phys.}
  {\bfseries 390} no.~1, (2022) 425--470},
  \href{http://arxiv.org/abs/2103.15826}{{\ttfamily arXiv:2103.15826
  [hep-th]}}.

\bibitem{Ashwinkumar:2021kav}
M.~Ashwinkumar, M.~Dodelson, A.~Kidambi, J.~M. Leedom, and M.~Yamazaki,
  ``{Chern-Simons invariants from ensemble averages},''
  \href{http://dx.doi.org/10.1007/JHEP08(2021)044}{{\em JHEP} {\bfseries 08}
  (2021) 044}, \href{http://arxiv.org/abs/2104.14710}{{\ttfamily
  arXiv:2104.14710 [hep-th]}}.

\bibitem{Dong:2021wot}
J.~Dong, T.~Hartman, and Y.~Jiang, ``{Averaging over moduli in deformed WZW
  models},'' \href{http://dx.doi.org/10.1007/JHEP09(2021)185}{{\em JHEP}
  {\bfseries 09} (2021) 185}, \href{http://arxiv.org/abs/2105.12594}{{\ttfamily
  arXiv:2105.12594 [hep-th]}}.

\bibitem{Collier:2021rsn}
S.~Collier and A.~Maloney, ``{Wormholes and spectral statistics in the Narain
  ensemble},'' \href{http://dx.doi.org/10.1007/JHEP03(2022)004}{{\em JHEP}
  {\bfseries 03} (2022) 004}, \href{http://arxiv.org/abs/2106.12760}{{\ttfamily
  arXiv:2106.12760 [hep-th]}}.

\bibitem{Benjamin:2021ygh}
N.~Benjamin, S.~Collier, A.~L. Fitzpatrick, A.~Maloney, and E.~Perlmutter,
  ``{Harmonic analysis of 2d CFT partition functions},''
  \href{http://dx.doi.org/10.1007/JHEP09(2021)174}{{\em JHEP} {\bfseries 09}
  (2021) 174}, \href{http://arxiv.org/abs/2107.10744}{{\ttfamily
  arXiv:2107.10744 [hep-th]}}.

\bibitem{Dymarsky:2021xfc}
A.~Dymarsky and A.~Sharon, ``{Non-rational Narain CFTs from codes over
  F$_{4}$},'' \href{http://dx.doi.org/10.1007/JHEP11(2021)016}{{\em JHEP}
  {\bfseries 11} (2021) 016}, \href{http://arxiv.org/abs/2107.02816}{{\ttfamily
  arXiv:2107.02816 [hep-th]}}.

\bibitem{Chakraborty:2021gzh}
S.~Chakraborty and A.~Hashimoto, ``{Weighted average over the Narain moduli
  space as a TTbar deformation of the CFT target space},''
  \href{http://dx.doi.org/10.1103/PhysRevD.105.086018}{{\em Phys. Rev. D}
  {\bfseries 105} no.~8, (2022) 086018},
  \href{http://arxiv.org/abs/2109.10382}{{\ttfamily arXiv:2109.10382
  [hep-th]}}.

\bibitem{Henriksson:2022dnu}
J.~Henriksson, A.~Kakkar, and B.~McPeak, ``{Narain CFTs and quantum codes at
  higher genus},'' \href{http://dx.doi.org/10.1007/JHEP04(2023)011}{{\em JHEP}
  {\bfseries 04} (2023) 011}, \href{http://arxiv.org/abs/2205.00025}{{\ttfamily
  arXiv:2205.00025 [hep-th]}}.

\bibitem{Kawabata:2022jxt}
K.~Kawabata, T.~Nishioka, and T.~Okuda, ``{Narain CFTs from qudit stabilizer
  codes},'' \href{http://dx.doi.org/10.21468/SciPostPhysCore.6.2.035}{{\em
  SciPost Phys. Core} {\bfseries 6} (2023) 035},
  \href{http://arxiv.org/abs/2212.07089}{{\ttfamily arXiv:2212.07089
  [hep-th]}}.

\bibitem{Kames-King:2023fpa}
J.~Kames-King, A.~Kanargias, B.~Knighton, and M.~Usatyuk, ``{The Lion, the
  Witch, and the Wormhole: Ensemble averaging the symmetric product
  orbifold},'' \href{http://arxiv.org/abs/2306.07321}{{\ttfamily
  arXiv:2306.07321 [hep-th]}}.

\bibitem{Alam:2023qac}
Y.~F. Alam, K.~Kawabata, T.~Nishioka, T.~Okuda, and S.~Yahagi, ``{Narain CFTs
  from nonbinary stabilizer codes},''
  \href{http://arxiv.org/abs/2307.10581}{{\ttfamily arXiv:2307.10581
  [hep-th]}}.

\bibitem{Kawabata:2023usr}
K.~Kawabata, T.~Nishioka, and T.~Okuda, ``{Supersymmetric conformal field
  theories from quantum stabilizer codes},''
  \href{http://dx.doi.org/10.1103/PhysRevD.108.L081901}{{\em Phys. Rev. D}
  {\bfseries 108} no.~8, (2023) L081901},
  \href{http://arxiv.org/abs/2307.14602}{{\ttfamily arXiv:2307.14602
  [hep-th]}}.

\bibitem{Kawabata:2023iss}
K.~Kawabata, T.~Nishioka, and T.~Okuda, ``{Narain CFTs from quantum codes and
  their $\mathbb{Z}_2$ gauging},''
  \href{http://arxiv.org/abs/2308.01579}{{\ttfamily arXiv:2308.01579
  [hep-th]}}.

\bibitem{Aharony:2023zit}
O.~Aharony, A.~Dymarsky, and A.~D. Shapere, ``{Holographic description of
  Narain CFTs and their code-based ensembles},''
  \href{http://arxiv.org/abs/2310.06012}{{\ttfamily arXiv:2310.06012
  [hep-th]}}.

\bibitem{Ashwinkumar:2023ctt}
M.~Ashwinkumar, A.~Kidambi, J.~M. Leedom, and M.~Yamazaki, ``{Generalized
  Narain Theories Decoded: Discussions on Eisenstein series, Characteristics,
  Orbifolds, Discriminants and Ensembles in any Dimension},''
  \href{http://arxiv.org/abs/2311.00699}{{\ttfamily arXiv:2311.00699
  [hep-th]}}.

\bibitem{Meruliya:2021utr}
V.~Meruliya, S.~Mukhi, and P.~Singh, ``{Poincar\'e Series, 3d Gravity and
  Averages of Rational CFT},''
  \href{http://dx.doi.org/10.1007/JHEP04(2021)267}{{\em JHEP} {\bfseries 04}
  (2021) 267}, \href{http://arxiv.org/abs/2102.03136}{{\ttfamily
  arXiv:2102.03136 [hep-th]}}.

\bibitem{Castro:2011zq}
A.~Castro, M.~R. Gaberdiel, T.~Hartman, A.~Maloney, and R.~Volpato, ``{The
  Gravity Dual of the Ising Model},''
  \href{http://dx.doi.org/10.1103/PhysRevD.85.024032}{{\em Phys. Rev. D}
  {\bfseries 85} (2012) 024032},
  \href{http://arxiv.org/abs/1111.1987}{{\ttfamily arXiv:1111.1987 [hep-th]}}.

\bibitem{Meruliya:2021lul}
V.~Meruliya and S.~Mukhi, ``{AdS$_{3}$ gravity and RCFT ensembles with multiple
  invariants},'' \href{http://dx.doi.org/10.1007/JHEP08(2021)098}{{\em JHEP}
  {\bfseries 08} (2021) 098}, \href{http://arxiv.org/abs/2104.10178}{{\ttfamily
  arXiv:2104.10178 [hep-th]}}.

\bibitem{Romaidis:2023zpx}
I.~Romaidis and I.~Runkel, ``{CFT correlators and mapping class group
  averages},'' \href{http://arxiv.org/abs/2309.14000}{{\ttfamily
  arXiv:2309.14000 [hep-th]}}.

\bibitem{Henriksson:2021qkt}
J.~Henriksson, A.~Kakkar, and B.~McPeak, ``{Classical codes and chiral CFTs at
  higher genus},'' \href{http://dx.doi.org/10.1007/JHEP05(2022)159}{{\em JHEP}
  {\bfseries 05} (2022) 159}, \href{http://arxiv.org/abs/2112.05168}{{\ttfamily
  arXiv:2112.05168 [hep-th]}}.

\bibitem{Buican:2021uyp}
M.~Buican, A.~Dymarsky, and R.~Radhakrishnan, ``{Quantum codes, CFTs, and
  defects},'' \href{http://dx.doi.org/10.1007/JHEP03(2023)017}{{\em JHEP}
  {\bfseries 03} (2023) 017}, \href{http://arxiv.org/abs/2112.12162}{{\ttfamily
  arXiv:2112.12162 [hep-th]}}.

\bibitem{Benini:2022hzx}
F.~Benini, C.~Copetti, and L.~Di~Pietro, ``{Factorization and global symmetries
  in holography},'' \href{http://dx.doi.org/10.21468/SciPostPhys.14.2.019}{{\em
  SciPost Phys.} {\bfseries 14} no.~2, (2023) 019},
  \href{http://arxiv.org/abs/2203.09537}{{\ttfamily arXiv:2203.09537
  [hep-th]}}.

\bibitem{Henriksson:2022dml}
J.~Henriksson and B.~McPeak, ``{Averaging over codes and an SU(2) modular
  bootstrap},'' \href{http://dx.doi.org/10.1007/JHEP11(2023)035}{{\em JHEP}
  {\bfseries 11} (2023) 035}, \href{http://arxiv.org/abs/2208.14457}{{\ttfamily
  arXiv:2208.14457 [hep-th]}}.

\bibitem{Raeymaekers:2021ypf}
J.~Raeymaekers, ``{A note on ensemble holography for rational tori},''
  \href{http://dx.doi.org/10.1007/JHEP12(2021)177}{{\em JHEP} {\bfseries 12}
  (2021) 177}, \href{http://arxiv.org/abs/2110.08833}{{\ttfamily
  arXiv:2110.08833 [hep-th]}}.

\bibitem{Cotler:2020hgz}
J.~Cotler and K.~Jensen, ``{AdS$_3$ wormholes from a modular bootstrap},''
  \href{http://dx.doi.org/10.1007/JHEP11(2020)058}{{\em JHEP} {\bfseries 11}
  (2020) 058}, \href{http://arxiv.org/abs/2007.15653}{{\ttfamily
  arXiv:2007.15653 [hep-th]}}.

\bibitem{Elitzur:1989nr}
S.~Elitzur, G.~W. Moore, A.~Schwimmer, and N.~Seiberg, ``{Remarks on the
  Canonical Quantization of the Chern-Simons-Witten Theory},''
  \href{http://dx.doi.org/10.1016/0550-3213(89)90436-7}{{\em Nucl. Phys. B}
  {\bfseries 326} (1989) 108--134}.

\bibitem{Cotler:2018zff}
J.~Cotler and K.~Jensen, ``{A theory of reparameterizations for AdS$_3$
  gravity},'' \href{http://dx.doi.org/10.1007/JHEP02(2019)079}{{\em JHEP}
  {\bfseries 02} (2019) 079}, \href{http://arxiv.org/abs/1808.03263}{{\ttfamily
  arXiv:1808.03263 [hep-th]}}.

\bibitem{Kapustin:2010if}
A.~Kapustin and N.~Saulina, ``{Surface operators in 3d Topological Field Theory
  and 2d Rational Conformal Field Theory},''
  \href{http://arxiv.org/abs/1012.0911}{{\ttfamily arXiv:1012.0911 [hep-th]}}.

\bibitem{Fuchs:2002cm}
J.~Fuchs, I.~Runkel, and C.~Schweigert, ``{TFT construction of RCFT correlators
  1. Partition functions},''
  \href{http://dx.doi.org/10.1016/S0550-3213(02)00744-7}{{\em Nucl. Phys. B}
  {\bfseries 646} (2002) 353--497},
  \href{http://arxiv.org/abs/hep-th/0204148}{{\ttfamily arXiv:hep-th/0204148}}.

\bibitem{Choudhury:2021nal}
S.~Choudhury and K.~Shirish, ``{Wormhole calculus without averaging from
  O(N)q-1 tensor model},''
  \href{http://dx.doi.org/10.1103/PhysRevD.105.046002}{{\em Phys. Rev. D}
  {\bfseries 105} no.~4, (2022) 046002},
  \href{http://arxiv.org/abs/2106.14886}{{\ttfamily arXiv:2106.14886
  [hep-th]}}.

\bibitem{Heckman:2021vzx}
J.~J. Heckman, A.~P. Turner, and X.~Yu, ``{Disorder averaging and its UV
  discontents},'' \href{http://dx.doi.org/10.1103/PhysRevD.105.086021}{{\em
  Phys. Rev. D} {\bfseries 105} no.~8, (2022) 086021},
  \href{http://arxiv.org/abs/2111.06404}{{\ttfamily arXiv:2111.06404
  [hep-th]}}.

\bibitem{Baume:2023kkf}
F.~Baume, J.~J. Heckman, M.~H\"ubner, E.~Torres, A.~P. Turner, and X.~Yu,
  ``{SymTrees and Multi-Sector QFTs},''
  \href{http://arxiv.org/abs/2310.12980}{{\ttfamily arXiv:2310.12980
  [hep-th]}}.

\bibitem{Moore:1989yh}
G.~W. Moore and N.~Seiberg, ``{Taming the Conformal Zoo},''
  \href{http://dx.doi.org/10.1016/0370-2693(89)90897-6}{{\em Phys. Lett. B}
  {\bfseries 220} (1989) 422--430}.

\bibitem{Moore:1988ss}
G.~W. Moore and N.~Seiberg, ``{Naturality in Conformal Field Theory},''
  \href{http://dx.doi.org/10.1016/0550-3213(89)90511-7}{{\em Nucl. Phys. B}
  {\bfseries 313} (1989) 16--40}.

\bibitem{Maldacena:1998bw}
J.~M. Maldacena and A.~Strominger, ``{AdS(3) black holes and a stringy
  exclusion principle},''
  \href{http://dx.doi.org/10.1088/1126-6708/1998/12/005}{{\em JHEP} {\bfseries
  12} (1998) 005}, \href{http://arxiv.org/abs/hep-th/9804085}{{\ttfamily
  arXiv:hep-th/9804085}}.

\bibitem{Polchinski:1998rq}
J.~Polchinski, \href{http://dx.doi.org/10.1017/CBO9780511816079}{{\em {String
  theory. Vol. 1: An introduction to the bosonic string}}}.
\newblock Cambridge Monographs on Mathematical Physics. Cambridge University
  Press, 12, 2007.

\bibitem{DiFrancesco:1997nk}
P.~Di~Francesco, P.~Mathieu, and D.~Senechal,
  \href{http://dx.doi.org/10.1007/978-1-4612-2256-9}{{\em {Conformal Field
  Theory}}}.
\newblock Graduate Texts in Contemporary Physics. Springer-Verlag, New York,
  1997.

\bibitem{Kraus:2006nb}
P.~Kraus and F.~Larsen, ``{Partition functions and elliptic genera from
  supergravity},'' \href{http://dx.doi.org/10.1088/1126-6708/2007/01/002}{{\em
  JHEP} {\bfseries 01} (2007) 002},
  \href{http://arxiv.org/abs/hep-th/0607138}{{\ttfamily arXiv:hep-th/0607138}}.

\bibitem{Cappelli:1986hf}
A.~Cappelli, C.~Itzykson, and J.~B. Zuber, ``{Modular Invariant Partition
  Functions in Two-Dimensions},''
  \href{http://dx.doi.org/10.1016/0550-3213(87)90155-6}{{\em Nucl. Phys. B}
  {\bfseries 280} (1987) 445--465}.

\bibitem{Cappelli:1987xt}
A.~Cappelli, C.~Itzykson, and J.~B. Zuber, ``{The ADE Classification of Minimal
  and A1(1) Conformal Invariant Theories},''
  \href{http://dx.doi.org/10.1007/BF01221394}{{\em Commun. Math. Phys.}
  {\bfseries 113} (1987) 1}.

\bibitem{Omami:2009}
R.~Omami, M.~Omami, and R.~Ouni, ``Group of square roots of unity modulo n,''
  \href{https://publications.waset.org/vol/31}{{\em International Journal of
  Mathematical and Computational Sciences} {\bfseries 3} no.~7, (2009) 505 --
  513}.

\bibitem{Kapustin:2010hk}
A.~Kapustin and N.~Saulina, ``{Topological boundary conditions in abelian
  Chern-Simons theory},''
  \href{http://dx.doi.org/10.1016/j.nuclphysb.2010.12.017}{{\em Nucl. Phys. B}
  {\bfseries 845} (2011) 393--435},
  \href{http://arxiv.org/abs/1008.0654}{{\ttfamily arXiv:1008.0654 [hep-th]}}.

\bibitem{Roumpedakis_2023}
K.~Roumpedakis, S.~Seifnashri, and S.-H. Shao, ``Higher gauging and
  non-invertible condensation defects,''
  \href{https://doi.org/10.1007%2Fs00220-023-04706-9}{{\em Communications in
  Mathematical Physics} {\bfseries 401} no.~3, (May, 2023) 3043--3107}.

\bibitem{Eberhardt:2021jvj}
L.~Eberhardt, ``{Summing over Geometries in String Theory},''
  \href{http://dx.doi.org/10.1007/JHEP05(2021)233}{{\em JHEP} {\bfseries 05}
  (2021) 233}, \href{http://arxiv.org/abs/2102.12355}{{\ttfamily
  arXiv:2102.12355 [hep-th]}}.

\bibitem{Maldacena:2004rf}
J.~M. Maldacena and L.~Maoz, ``{Wormholes in AdS},''
  \href{http://dx.doi.org/10.1088/1126-6708/2004/02/053}{{\em JHEP} {\bfseries
  02} (2004) 053}, \href{http://arxiv.org/abs/hep-th/0401024}{{\ttfamily
  arXiv:hep-th/0401024}}.

\bibitem{Paolo}
w.~i.~p. Paolo~Rossi.

\bibitem{Furuta:2023xwl}
Y.~Furuta, ``{On the Rationality and the Code Structure of a Narain CFT, and
  the Simple Current Orbifold},''
  \href{http://arxiv.org/abs/2307.04190}{{\ttfamily arXiv:2307.04190
  [hep-th]}}.

\bibitem{Porrati:2019knx}
M.~Porrati and C.~Yu, ``{Kac-Moody and Virasoro Characters from the
  Perturbative Chern-Simons Path Integral},''
  \href{http://dx.doi.org/10.1007/JHEP05(2019)083}{{\em JHEP} {\bfseries 05}
  (2019) 083}, \href{http://arxiv.org/abs/1903.05100}{{\ttfamily
  arXiv:1903.05100 [hep-th]}}.

\bibitem{Porrati:2021sdc}
M.~Porrati and C.~Yu, ``{Partition functions of Chern-Simons theory on
  handlebodies by radial quantization},''
  \href{http://dx.doi.org/10.1007/JHEP07(2021)194}{{\em JHEP} {\bfseries 07}
  (2021) 194}, \href{http://arxiv.org/abs/2104.12799}{{\ttfamily
  arXiv:2104.12799 [hep-th]}}.

\bibitem{Raeymaekers:2020gtz}
J.~Raeymaekers, ``{Conical spaces, modular invariance and $c_{p,1}$
  holography},'' \href{http://dx.doi.org/10.1007/JHEP03(2021)189}{{\em JHEP}
  {\bfseries 03} (2021) 189}, \href{http://arxiv.org/abs/2012.07934}{{\ttfamily
  arXiv:2012.07934 [hep-th]}}.

\bibitem{Campoleoni:2017xyl}
A.~Campoleoni, S.~Fredenhagen, and J.~Raeymaekers, ``{Quantizing higher-spin
  gravity in free-field variables},''
  \href{http://dx.doi.org/10.1007/JHEP02(2018)126}{{\em JHEP} {\bfseries 02}
  (2018) 126},
\href{http://arxiv.org/abs/1712.08078}{{\ttfamily arXiv:1712.08078 [hep-th]}}.
%%CITATION = ARXIV:1712.08078;%%.

\end{thebibliography}\endgroup

\end{document}